\documentclass[preprint,12pt]{elsarticle}

\usepackage{graphicx}
\usepackage{subfigure}
\usepackage{amssymb}
\usepackage{multirow}
\usepackage{lineno}
\usepackage{url}
\usepackage{amsmath}
\usepackage[letterpaper,top=1.2in,bottom=1in,left=1.2in,right=1.2in]{geometry}%
\usepackage{color}

\usepackage{xspace}

\journal{}

\begin{document}

\begin{frontmatter}


\title{DeepM\&Mnet: Inferring the electroconvection multiphysics fields based on operator approximation by neural networks}



\author[Brown]{Shengze Cai\fnref{fn1}}
\author[Brown]{Zhicheng Wang\fnref{fn1}}
\author[MIT]{Lu Lu}
\author[JHU]{Tamer A. Zaki}
\author[Brown]{George Em Karniadakis\corref{cor1}}
\cortext[cor1]{Corresponding author: }
\ead{george\_karniadakis@brown.edu}

\fntext[fn1]{The authors contributed equally to this work.}

\address[Brown]{Division of Applied Mathematics, Brown University, Providence, RI, 02912, USA}
\address[MIT]{Department of Mathematics, Massachusetts Institute of Technology, Cambridge, MA, 02139, USA}
\address[JHU]{Department of Mechanical Engineering, Johns Hopkins University, Baltimore, MD 21218, USA}


\begin{abstract}
Electroconvection is a multiphysics problem involving coupling of the flow field with the electric field as well as the cation and anion concentration fields. For small Debye lengths, very steep boundary layers are developed, but standard numerical methods can simulate the different regimes quite accurately. Here, we use electroconvection as a benchmark problem to put forward a new data assimilation framework, the DeepM\&Mnet, for simulating multiphysics and multiscale problems at speeds 
much faster than standard numerical methods using pre-trained neural networks (NNs). We first pre-train DeepONets that can predict independently each field, given general inputs from the rest of the fields of the coupled system. DeepONets can approximate nonlinear operators and are composed of two sub-networks, a {\em branch net} for the input fields and a {\em trunk net} for the locations of the output field. DeepONets, which are extremely fast, are used as building blocks in the  DeepM\&Mnet and form constraints for the multiphysics solution along with some sparse available measurements of any of the fields. We demonstrate the new methodology and  document the accuracy of each individual DeepONet, and subsequently we present two different DeepM\&Mnet architectures  that infer accurately and efficiently 2D electroconvection fields for unseen electric potentials. The DeepM\&Mnet framework is general and can be applied for building any complex multiphysics and multiscale models based on very few measurements using pre-trained DeepONets in a ``plug-and-play" mode. 
\end{abstract}


\end{frontmatter}



\section{Introduction}\label{sec:intro} %

Recently, deep learning techniques have been introduced in modeling diverse fluid mechanics problems  \cite{brunton2020machine,duraisamy2019turbulence,Dixia2020}. A promising direction is focused on the surrogate modeling of complex fluid systems, which are typically described by partial differential equations (PDEs). For example, \citet{raissi2019physics} applied physics-informed neural networks (PINNs) to obtain accurate solutions of a wide class of PDEs. The governing equations are embedded in PINNs and form the constraints of the solutions. Similar approaches for scientific 
machine learning have been proposed in a number of recent works \cite{sirignano2018dgm,han2018solving,long2019pde}.
Following the framework of \cite{raissi2019physics}, a number of learning-based methods were proposed to simulate different types of fluid mechanics problems \cite{raissi2019deep,raissi2020hidden,sun2020surrogate,jin2020nsfnets}. 
In addition to solving forward problems, neural networks are also used to identify the unknown coefficient values of PDEs from data \cite{ling2016reynolds, zhang2019quantifying,pang2019fpinns,chen2020physics,Yazdani865063}. A typical example is to learn the Reynolds number of Navier-Stokes equations based on some measurements of flow \cite{rudy2017data, raissi2017physicsB}. This is a data-driven inverse problem, which is not easy to address by conventional computational fluid mechanics (CFD) methods. 

Overall, the recent applications of deep learning to physics modeling are based on the universal approximation theorem stating that neural networks (NNs) can be used to approximate any continuous function \cite{hornik1989multilayer}. However, there are other approximation theorems stating that a neural network can approximate accurately any continuous nonlinear functional \cite{chen1993approximations} or operator (a mapping from a function to another function) \cite{chen1995universal}. 
Based on the universal approximation theorem for operator, \citet{lu2019deeponet} proposed a specific network architecture, the deep operator network (DeepONet), to learn both explicit as well as implicit  operators, e.g., in the form of PDEs. DeepONet, which is designed based on rigorous theory, is composed of two sub-networks, the {\em branch net} for the input function and the {\em trunk net} for the coordinate to evaluate the output function. 
In \cite{lu2019deeponet}, the authors presented theoretical analysis and demonstrated various examples to show the high accuracy and high efficiency of DeepONets. 
Inspired by the work of \cite{lu2019deeponet}, here we first pose the related question:  
``Can DeepONet learn the mapping between two coupled state variables for complex fluid systems?"
If this is the case, we ask: ``Can the pre-trained DeepONets be combined in a simplified data assimilation framework, using a single NN, to accelerate significantly modeling of 
multiphysics and multiscale flow problems?"

In this paper, our first goal is to demonstrate the effectiveness of DeepONets for multiphysics and multiscale problems in fluid mechanics. To this end, we use electroconvection as a benchmark problem, which describes electrolyte flow driven by an electric voltage and involves multiphysics, i.e., the mass, momentum and cation/anion transport as well as electrostatic fields  \cite{mani2020electroconvection}. Electroconvection is observed in many electrochemical and microfluidic systems that usually have ion selective interfaces. Since the development of the theoretical formulations of electroconvection  by \cite{Rubinstein2000,Zaltzman2007}, which were later confirmed by experiment in \cite{Rubinstein2008}, a considerable number of numerical studies has been published \cite{Druzgalski2013PoF,Karatay2015,karatay2016coupling,mani2020electroconvection}. More recently, several 3D interesting simulations have been presented in  \cite{Druzgalski2016PRF,Pham2016PRE}. The main bottleneck of the simulation of electroconvection is the computational cost to resolve the electric double layer (EDL), the thickness of which is characterized by a dimensionless parameter, namely the Debye length ($\epsilon$). In the aforementioned simulations, $\epsilon=10^{-3}$ is used,  which is larger than the physical realistic value in experiments, i.e., $\epsilon<10^{-4}$, although studies show that the mean transport rate is only weakly sensitive 
to $\epsilon$ \cite{Druzgalski2016PRF}. Here, we aim to present a proof-of-concept
for a new deep learning framework and we use electronvection as a benchmark multiphysics problem.

To apply DeepONets to electroconvection, we train DeepONets to predict each field independently by giving inputs from the rest of the fields. Upon supervised training with labeled data, the DeepONets can predict very accurately and efficiently the target field. 
There are some existing learning-based methods that can perform similar input-output mappings. In particular, the most common algorithm is the convolutional neural networks (CNNs), which are used to learn mapping between two variables when the input and output functions are treated as images \cite{zhu2019physics,bhatnagar2019prediction,cai2019dense,kim2020prediction}. 
For example, \citet{kim2020prediction} used a CNN to predict the wall-normal heat flux from the wall-shear stresses and pressure. 
However, this kind of approach is limited to problems where the input and output are represented by 2D maps with equispaced grids. 
On the contrary, it is 
not necessary to include all the grid values inside the computational domain for the training data.

\begin{figure}[t]
\begin{center}
\includegraphics[width=\textwidth]{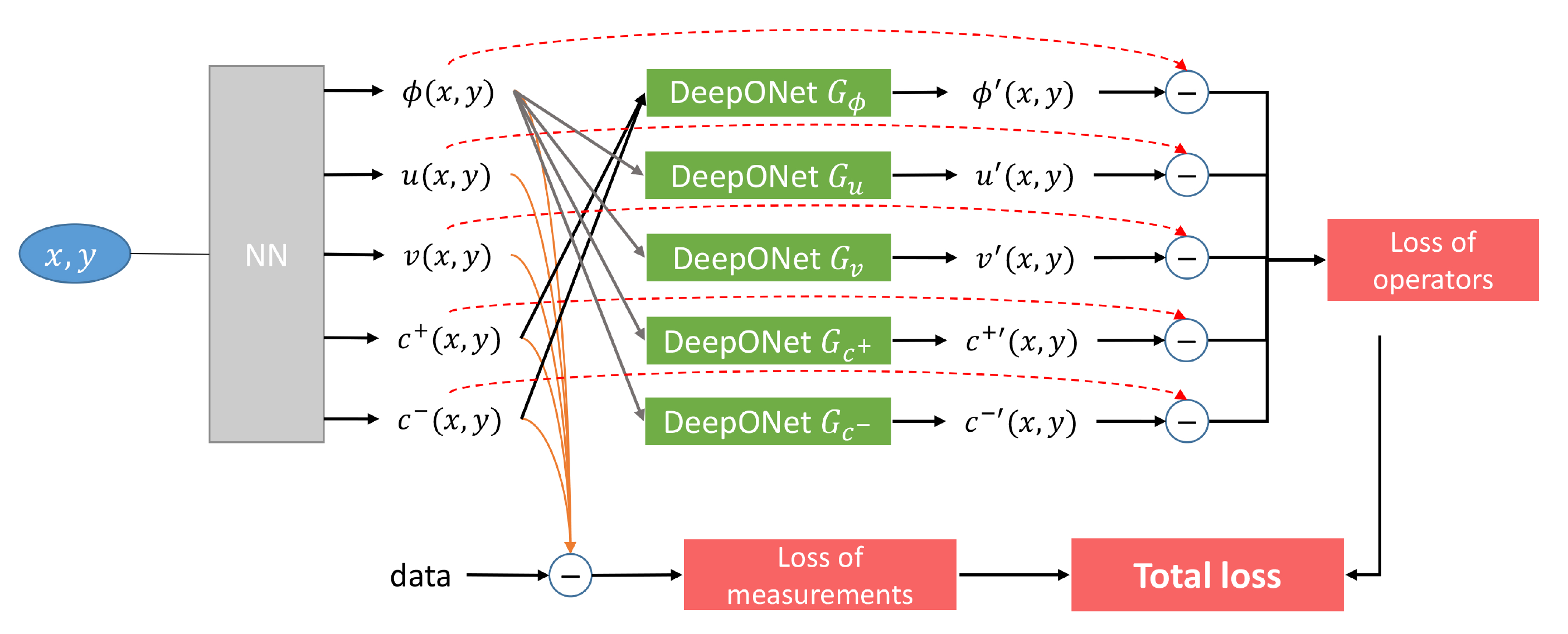}
\caption{DeepM\&Mnet for 2D electroconvection problem: schematic of the parallel architecture. Here, the neural network (NN), which is used to approximate the multiphysics solutions, is a fully-connected network with trainable parameters. The pre-trained DeepONets are fixed (not trainable) and they are considered as the constraints of the NN outputs. }
\label{fig:EC2D_DeepMM_parallel}
\end{center}
\end{figure}

In addition to the application of DeepONets, another objective in this paper is to put forward a new framework, the DeepM\&Mnet, for simulating general multiphysics and multiscale problems. To use the DeepONets for prediction of an independent condition, at least one of the electroconvection fields is required. However, this is not realistic in general. Therefore, we develop the DeepM\&Mnet, which can integrate several pre-trained DeepONets with a few measurements from any of the fields to produce the full fields of the coupled multiphysics system. To this end, we develop two architectures of DeepM\&Mnet. One of them, the parallel DeepM\&Mnet, is illustrated in Figure \ref{fig:EC2D_DeepMM_parallel}. 
In the context of DeepM\&Mnet, a neural network, which takes the spatial coordinates as inputs, is applied to approximate the multiphysics solutions. 
In other words, the neural network is the surrogate model of the multiphysics electroconvection. 
Then, the pre-trained DeepONets are used as building blocks and form the constraints for the solutions. The proposed DeepM\&Mnet can be considered as a framework assimilating the data measurements and the physics-related operators, but it is much more flexible and efficient than any other conventional numerical method in terms of dealing with such assimilation problem.

The paper is organized as follows. In Section \ref{sec:simulation}, we introduce the numerical simulation of the electroconvection problem and the data preparation. In Section \ref{sec:DeepONets}, we introduce the concept of DeepONet and train several DeepONets to predict separately each field accurately. In Section \ref{sec:DeepMM}, we propose two architectures of DeepM\&Mnet, namely the parallel DeepM\&Mnet and series DeepM\&Mnet, and then evaluate the performance of these new  
networks. Finally, we conclude the paper in Section \ref{sec:conclusion}. We also present some details on the implementations of DeepONet and DeepM\&Mnet, e.g., the choice of the training loss, regularization issues, etc., in the appendices.

\section{Numerical simulation of electroconvection and data generation}\label{sec:simulation} %




We developed an eletroconvection solver using the high-order spectral element method \cite{karniadakis2013spectral}. Following the numerical study in \cite{Druzgalski2013PoF}, here the governing equations in dimensionless form, including the Stokes equations, the electric potential and the ion transport equations, are given as follows:
\begin{subequations}
\begin{align}
0    &= -  \nabla p + \nabla^{2}\mathbf{u} + \mathbf{f_{e}}, \\
\nabla \cdot \mathbf{u} &= 0,  \\
-2\epsilon^{2}\nabla^{2}\phi &= \rho_{e}, \\
\frac{\partial c^{\pm}}{\partial t} &= - \nabla \cdot \mathbf{J^{\pm}},  \label{eq:EC_d_2D}  \\ 
\mathbf{J^{\pm}} &= c^{\pm}\mathbf{u}-\nabla  c^{\pm} \mp c^{\pm} \nabla \phi.  \label{eq:EC_e_2D}
\end{align}
\label{eq:electroConvection_2D}
\end{subequations}
Here, $\mathbf{u}$ is the velocity vector, $p$ is the pressure field and $\phi$ is the electric potential. Moreover, $c^{+}$ and $c^{-}$ are the cation and anion concentrations, respectively. Also, $\rho_{e}=z(c^{+}-c^{-})$ is the free
charge density with ionic valence $z=\pm1$, $f_{e}=-\kappa\rho_{e}\nabla\phi/2\epsilon^{2}$ is the electrostatic body force, where $\epsilon$ is the Debye length; $\kappa$ is the electrohydrodynamic coupling constant and its value is 0.5 throughout this paper. 

\begin{figure}[ht]
\begin{center}
\includegraphics[width=0.75\textwidth,trim=10 200 10 150,clip]{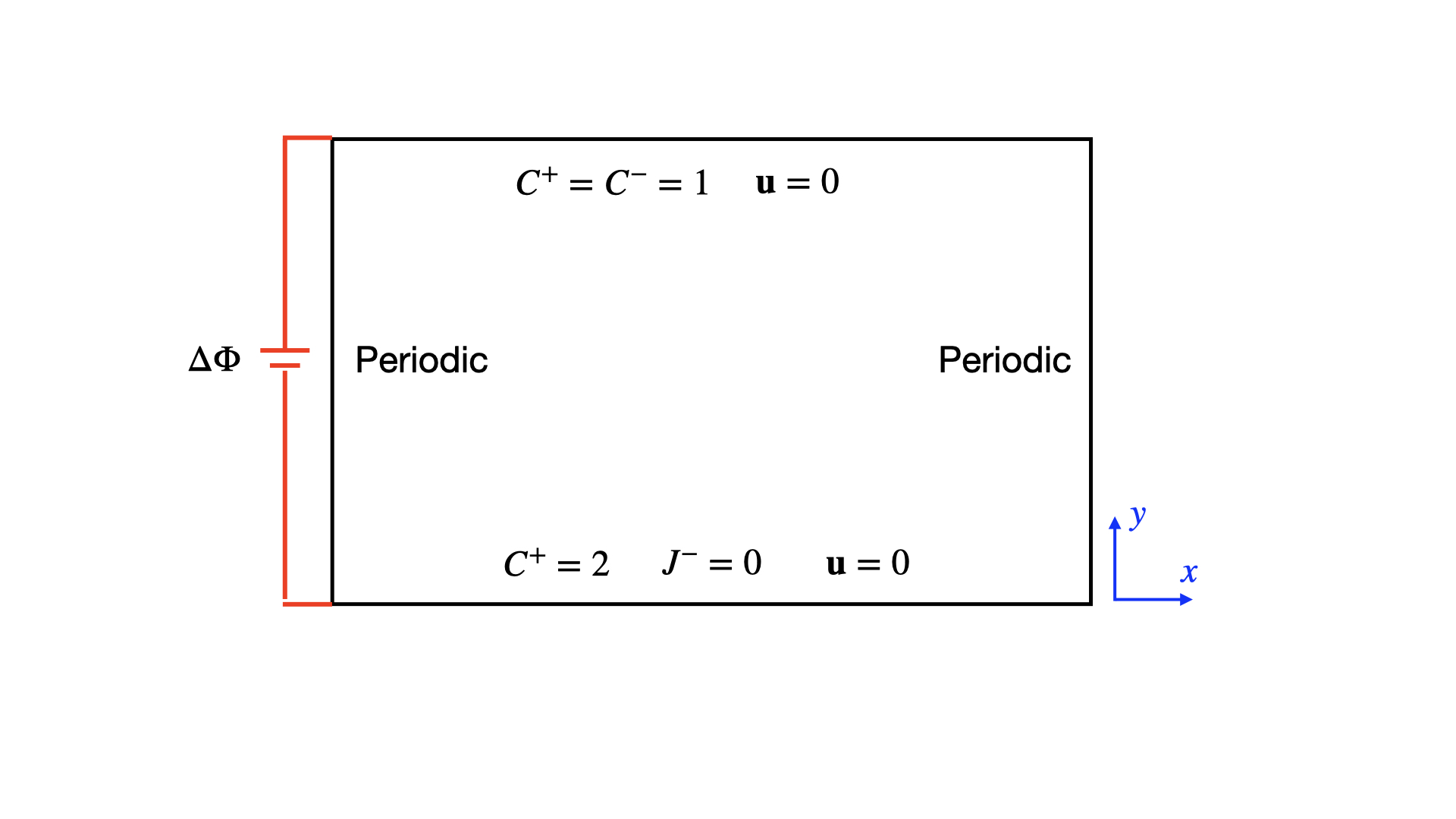}
\caption{Simulation domain and boundary conditions for 2D electroconvection.}\label{fig:bc}
\end{center}
\end{figure}

We consider electroconvection between an ion-selective surface and a stationary reservoir subject to an applied voltage $\Delta \Phi$. As shown in the Figure \ref{fig:bc}, the simulation domain is a rectangle of size $[-3,3]\times[0,1]$. On the top surface ($y=1$), no-slip and no-penetration wall boundary condition ($\mathbf{u}=0$) for the Stokes equations, constant ion concentrations ($c^{+}=c^{-}=1$) and constant potential ($\phi=\Delta \Phi$) are imposed. The bottom surface ($y=0$) is also a solid wall, and it is assumed to be impermeable to anions ($J^{-}=0$), but permeable to cations whose concentration is maintained at $c^{+}=2$. The potential on the bottom surface is $\phi=0$. Periodic boundary conditions are used on both the right ($x=3$) and left ($x=-3$) boundaries. The electroconvection solver is developed based on our spectral element library NekTar. The spectral element is used to discretize the governing equations in space and the second-order stiffly-stable scheme is employed for time advancing \cite{karniadakis2013spectral}.

\begin{figure}[ht]
\centering
 \subfigure[]{      \includegraphics[width=0.475\textwidth,trim=50 0 60 0,clip]{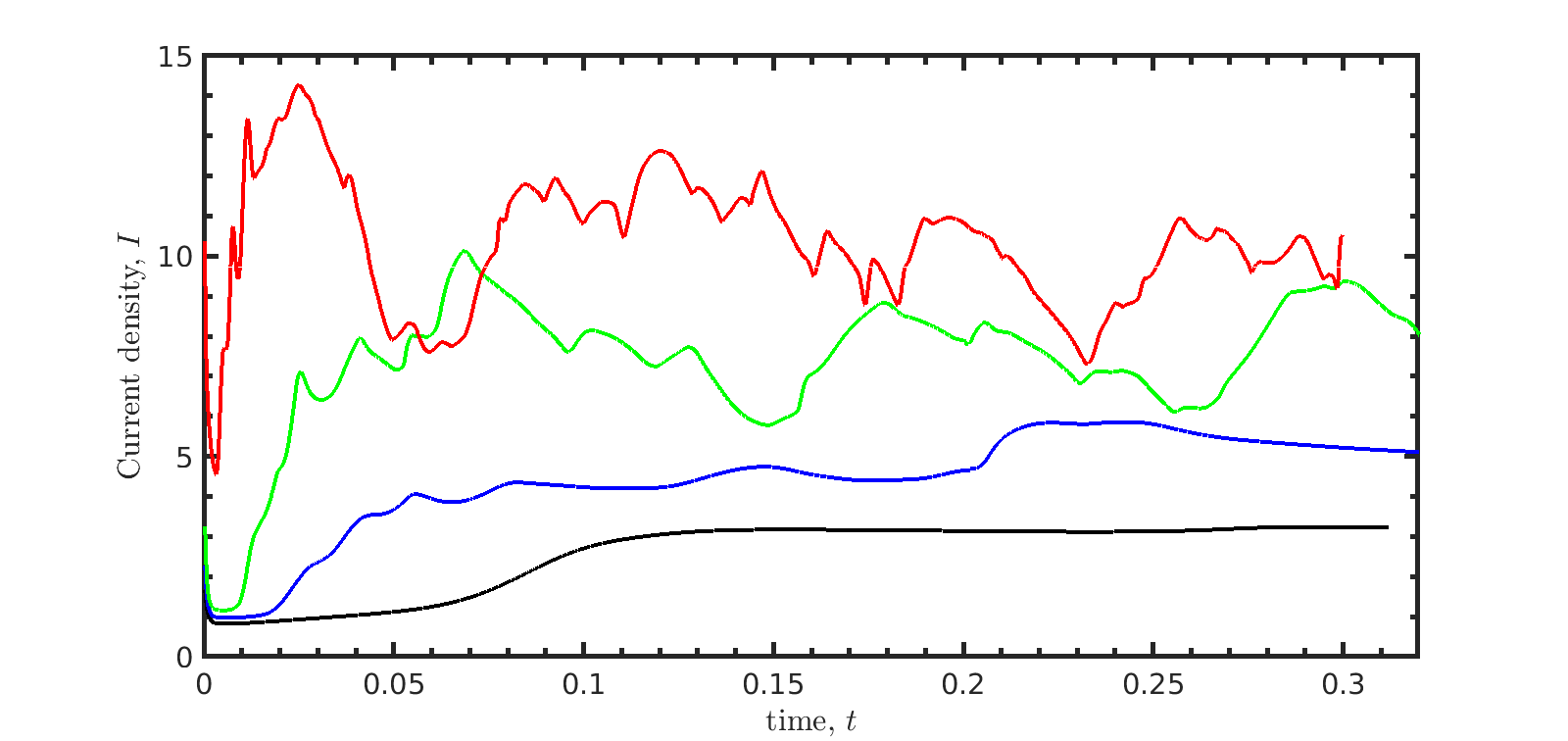}}
~
 \subfigure[]{
      \includegraphics[width=0.475\textwidth,trim=50 0 60 10,clip]{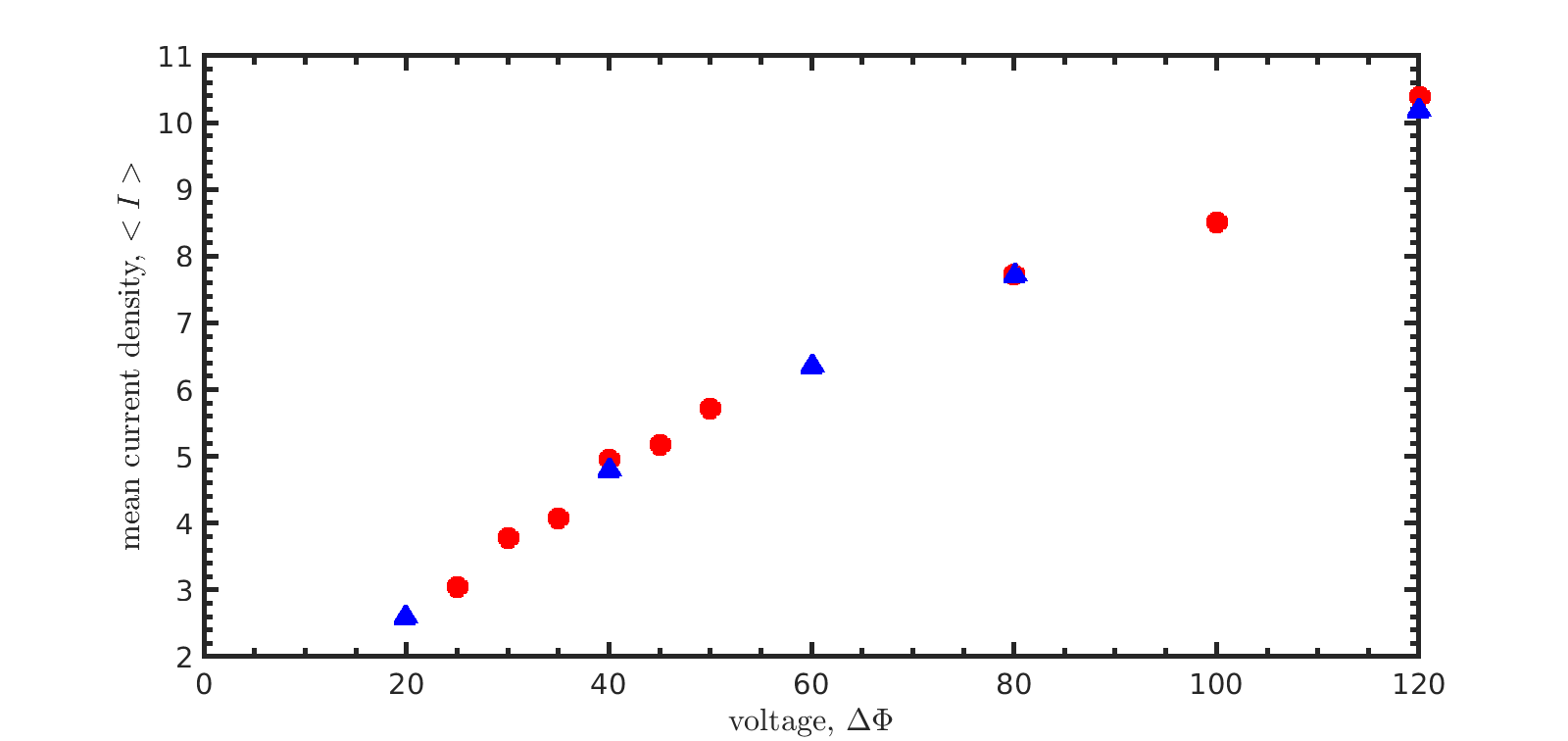}}
\caption{Spectral element simulation results of the current density ($I$) on the top wall and  validation against the simulations of \cite{Druzgalski2013PoF}. (a) $x$-averaged $I$  as a function of time ($t$): black line: applied voltage $\Delta \Phi=25$; blue line: $\Delta \Phi=40$; green line: $\Delta \Phi=80$; red line: $\Delta \Phi= 120$. (b) Time and $x$-averaged $<I>$ as a function of $\Delta \Phi$: red circles: NekTar simulation; blue triangles: simulation results of  \cite{Druzgalski2013PoF}.
}
\label{fig:EC2D_simulation}

\end{figure}

The key parameters for the eletroconvection system are $\epsilon$ and $\Delta \Phi$. In this paper, in order to validate the spectral element electroconvection solver, we first performed nine cases of 2D simulations at a fixed $\epsilon=10^{-3}$, while $\Delta \Phi=25$, 30, 35, 40, 50, 55, 80, 100 and 120, respectively. The computational domain consists of $48\times 64$ quadrilateral elements, which are uniformly distributed in the $x$ direction, but clustered in the vicinity of the walls. The smallest element size in the $y$ direction is $0.0011$. Same as the procedure in \cite{Druzgalski2013PoF}, the initial condition is generated by solving the governing equations without flow, and randomly perturbing the resulting concentration fields locally by 1\%. Moreover, the time step is $\delta t=5\times 10^{-7}$, and 4 spectral element modes ($3^{rd}$ order polynomials) and 7 quadrature points (per direction) are used in all the validation cases. All the simulations stopped after $t > 0.3$ and convergence was verified. The collection of the statistical samples began at $t=0.15$ when the current on the top surface entered into a statistically stationary state. Figure \ref{fig:EC2D_simulation} shows the comparisons of current spectral element solutions to those of \cite{Druzgalski2013PoF}. Same as the finding in \cite{Druzgalski2013PoF}, it can be observed from Figure \ref{fig:EC2D_simulation} (a) that the current ($I$) on the top surface is steady at low voltage $\Delta \Phi=25$, it becomes unsteady when $\Delta \Phi=40$, and it shows chaotic behaviour as the $\Delta \Phi$ is increased further. Quantitatively, the time averaged current $<I>$ is in good agreement with the results of \cite{Druzgalski2013PoF} , as shown in Figure \ref{fig:EC2D_simulation} (b).

With the spectral element code being fully validated, subsequently in order to generate the training data we performed simulations at $\epsilon=10^{-2}$ with $\Delta \Phi$ varying in the range $[5,75]$ systematically. 
The flow is steady under these conditions. Note that the computational domain is the same as that used in the validation, but the number of elements is reduced to $32\times 32$. Also since a bigger $\epsilon$ is used, the smallest element size in $y$ direction is increased to $0.0082$, the time step is increased to $\delta t=10^{-5}$, the number of spectral element modes is 3 and the number of quadrature points in each element in each direction is 5. All the simulations stopped at $t=2.5$ when the time is sufficient long for the current on the top surface to reach steady state. For each simulation condition, we 
save the 2D snapshots of $\phi$, $u$, $v$ and $c^{\pm}$ at time instant $t=2.5$, which will be used as the training data.


\section{DeepONets for 2D Electroconvection}\label{sec:DeepONets}

\subsection{The building blocks: DeepONets} 

In this section, we describe the DeepONets that will be used as the building blocks in DeepM\&Mnets. The DeepONet was proposed in \cite{lu2019deeponet} for learning general nonlinear operators,
including different types of partial differential equations (PDEs). 
Let $G$ be an operator mapping from a space of functions to another space of functions. In this study, $G$ is represented by a DeepONet, which takes inputs composed of two components: a function $U$ and the location points $(x,y)$, and outputs $G(U)(x,y)$. 
According to the physics of electroconvection, we design five independent DeepONets, which are divided into two classes. 
The first one takes the electric potential $\phi(x,y)$ as the input and predicts the velocity vector field  ($u,v$) and concentration fields ($c^{+},c^{-}$), which are denoted as $G_{u}$, $G_{v}$, $G_{c^{+}}$ and $G_{c^{-}}$, respectively. 
The second one is used to predict $\phi$ by using the cation and anion concentrations, which is represented by $G_{\phi}$. 
The definitions of these DeepONets are given in Table \ref{tab:DeepONets_2D_definition}. We should mention that the pressure field of the electroconvection is not considered in this work. However, it should be easy to include the pressure in the context of DeepONets as well as DeepM\&Mnets.

\begin{figure}[ht]
\begin{center}
\includegraphics[width=\textwidth]{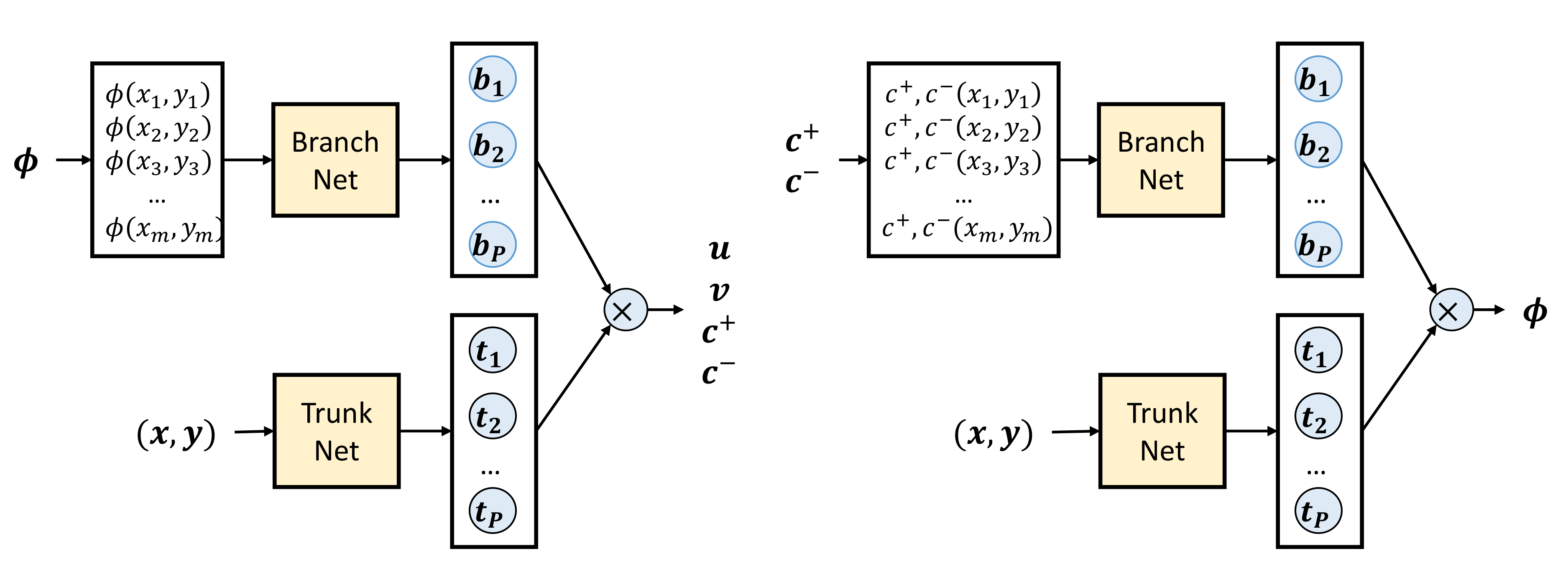}
\caption{DeepONets for 2D electroconvection: schematics of the DeepONet architectures. Left: using the electric potential to predict the velocity, cation and anion concentrations, i.e., $G_{u,v,c^{+},c^{-}}: \phi \rightarrow [u,v,c^{+},c^{-}]$; right: using the concentrations to predict the electric potential, i.e., $G_{\phi}: [c^{+},c^{-}] \rightarrow \phi$. 
}\label{fig:EC2D_DeepOnets_architecture}
\end{center}
\end{figure}

In this paper, we apply the ``unstacked" architecture for DeepONet, which is composed of a branch network and a trunk network. DeepONets are implemented in DeepXDE \cite{lu2019deepxde}, a user-friendly Python library designed for scientific machine learning. The schematic diagrams of the proposed networks are illustrated in Figure \ref{fig:EC2D_DeepOnets_architecture}. In this framework, 
the trunk network takes the coordinates $(x,y)$ as input and outputs $[t_{1},t_{2},\cdots,t_{p}]^{T} \in \mathbb{R}^{p} $. In addition, the input function, which is represented by $m$ discrete values (e.g., $[\phi(x_{1},y_{1}),\cdots,\phi(x_{m},y_{m})]^{T}$), is fed into the branch network. Then the two vectors from the branch and trunk nets are merged together via a dot product to obtain the output function value. We use the fully-connected networks as the sub-networks (i.e., trunk and branch nets) in this study. The numbers of hidden layers and the number of neurons per layer of these sub-networks are given in Table \ref{tab:DeepONets_2D_architecture}.

\begin{table}[th]
\begin{center}
\caption{DeepONets for 2D electroconvection: definitions of the 2D operators. 
 }\label{tab:DeepONets_2D_definition}
\begin{tabular}{|c|c|c|}
  \hline
  DeepONets & Input function & Output function \\
  \hline
  $G_{\phi}$  & $c^{+}(x,y),c^{-}(x,y)$    &  $\phi(x,y)$  \\
  $G_{u}$  & $\phi(x,y)$   &  $u(x,y)$  \\
  $G_{v}$  & $\phi(x,y)$   &  $v(x,y)$  \\
  $G_{c^{+}}$  & $\phi(x,y)$   &  $c^{+}(x,y)$  \\
  $G_{c^{-}}$  & $\phi(x,y)$   &  $c^{-}(x,y)$  \\
  \hline
\end{tabular}
\end{center}
\end{table}

\begin{table}[th]
\begin{center}
\caption{DeepONets for 2D electroconvection: architectures and loss functions of the DeepONets. MSE: mean squared error; MAPE: mean absolute percentage error. 
 }\label{tab:DeepONets_2D_architecture}
\begin{tabular}{|c|c|c|c|c|c|}
  \hline
  DeepONets & Branch depth & Branch width & Trunk depth & Trunk width & Loss \\
  \hline
  $G_{\phi}$  &  2  & $[300, 200]$  &  3   & $[100, 200, 200]$ & MSE \\
  $G_{u}$     &  2  & $[200, 200]$  &  3   & $[100, 200, 200]$ & MAPE\\
  $G_{v}$     &    2  & $[200, 200]$  &  3   & $[100, 200, 200]$  & MAPE \\
  $G_{c^{+}}$  &    2  & $[200, 200]$  &  3   & $[100, 200, 200]$ & MSE  \\
  $G_{c^{-}}$  &   2  & $[200, 200]$  &  3   & $[100, 200, 200]$  & MSE \\
  \hline
\end{tabular}
\end{center}
\end{table}

The DeepONets are trained by minimizing a loss function, which measures the difference between the labels and NN predictions. In general, the mean squared error (MSE) is applied:
\begin{equation}
\mathbf{MSE} = \frac{1}{N} \sum_{i=1}^{N}(V_{i} - \hat{V}_{i})^{2},
\label{eq:deepONet_MSE}
\end{equation}
where $V$ represents the predicting variable, $V_{i}$ and $\hat{V}_{i}$ are the labeled data and the prediction, respectively; $N$ is the number of training data. 
Alternatively, the mean absolute percentage error (MAPE) is also considered in the paper:
\begin{equation}
\mathbf{MAPE} = \frac{1}{N} \sum_{i=1}^{N} \frac{ |V_{i} - \hat{V}_{i} | }{ |V_{i}|+\eta },
\label{eq:deepONet_MAPE}
\end{equation}
where $\eta$ is a small number to guarantee the stability when $V_{i}=0$. The MSE loss works well in most cases, while the MAPE loss is better for the case where the output has a large range of function values. 
In this paper, we apply the MSE loss to the training of $G_{\phi,c^{+},c^{-}}$ and apply the MAPE loss for $G_{u,v}$. A comparison between MSE and MAPE applied to $G_{u}$ is demonstrated in \ref{sec:app_MSE_VS_MAPE}.

\subsection{Training of DeepONets}

\begin{figure}[t]
    \centering
    \subfigure[]{\label{fig:EC2D_DeepONets_dataDemo_2D}
    \includegraphics[width=0.95\textwidth]{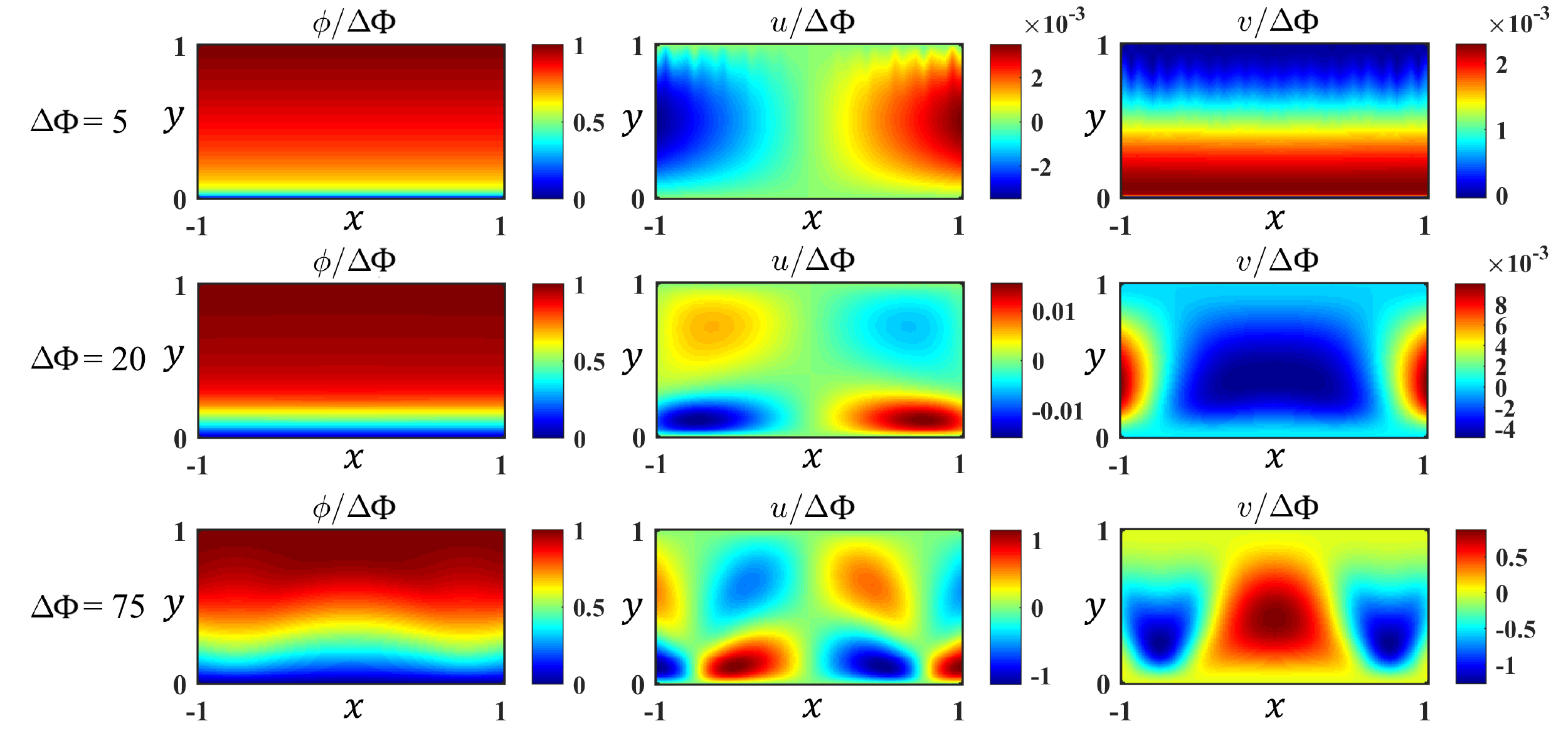}}
    \subfigure[]{\label{fig:EC2D_DeepONets_dataDemo_1D}
    \includegraphics[width=0.9\textwidth]{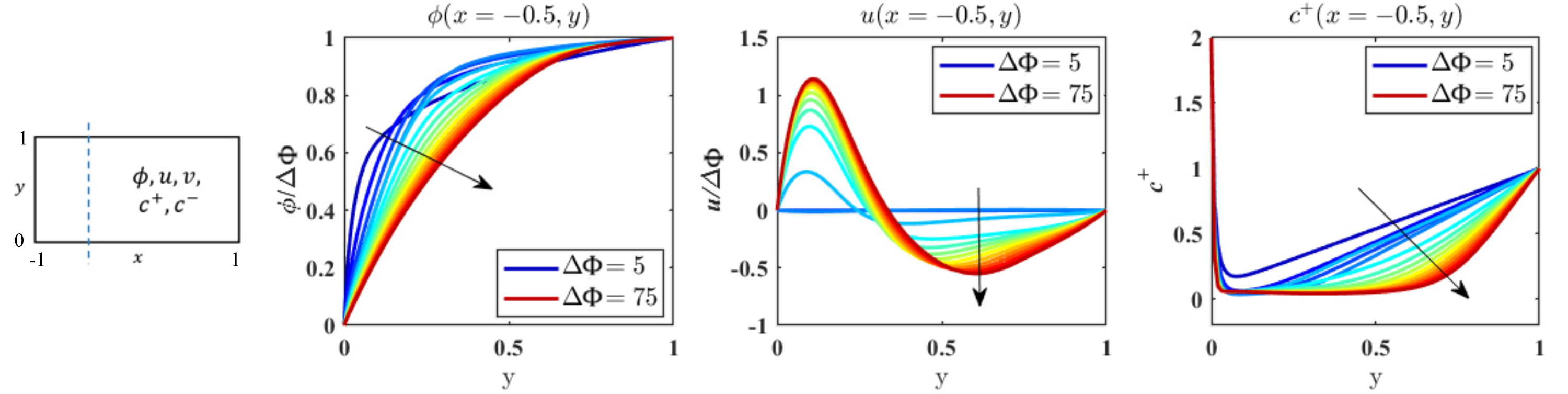}}
    \caption{DeepONets for 2D electroconvection: demonstration of the training data. (a) 2D steady fields of $\phi,u$ and $v$ under various boundary conditions, i.e., $\Delta\Phi=5, 20, 75$;
    (b) 1D profiles of $\phi(x=-0.5,y)$, $u(x=-0.5,y)$ and $c^{+}(x=-0.5,y)$ for  15 training states with $\Delta \Phi=5,10,\dots, 75$. The arrow represents the increasing direction of $\Delta \Phi$. 
    }\label{fig:EC2D_DeepONets_dataDemo}
\end{figure}

The training of DeepONets requires datasets with labeled outputs. 
We apply NekTar to simulate the 2D steady state fields of electroconvection problem. 
The computational domain is defined as $\Omega$: $x \in [-1,1]$, $y \in [0,1]$. 
As mentioned in Section \ref{sec:simulation}, different steady-state patterns can be produced by modifying the electric potential difference $\Delta \Phi$ between the boundaries.
The fields of $\phi(x,y)$, $c^{+}(x,y)$, $c^{-}(x,y)$ and velocities $u(x,y), v(x,y)$ are collected. By modifying the boundary conditions of $\phi$, namely using $\Delta \Phi=5,10,\dots, 75$, we generate 15 steady states for this electroconvection problem. 
The 2D snapshots at various values of $\Delta\Phi$ and some 1D profiles of $\phi(x=-0.5,y)$, $u(x=-0.5,y)$,  $c^{+}(x=-0.5,y)$ are demonstrated in Figure \ref{fig:EC2D_DeepONets_dataDemo}. 
Note that throughout this paper the data of $\phi$, $u$, $v$ are normalized by $\Delta \Phi$ (which is readily known) for enhanced stability in the  DeepONet training. From the figures, we can find the flow pattern varies  significantly with different $\Delta\Phi$. Moreover, the range of the velocity magnitude is very large ($10^{-4}-10^{0}$), showing the multiscale nature of this electroconvection problem.

For each 2D input field, we have $21\times11$ uniformly-distributed sensors to represent the function. As for the corresponding output fields, we randomly select 800 data points in the space for each state variable. In this context, we have $N=15\times800=12000$ training data points in all, where one data item is a triplet. For example, for the DeepONet $G_{u}$, one data item is $[\phi, (x,y), u(x,y)]$.  
We also use NekTar to generate fields under two additional conditions, namely $\Delta \Phi=13.4$ and $\Delta \Phi=62.15$, which are not included in the training datasets and are used for testing and validation. 
As mentioned in the introduction, the training data can be selected randomly in the computational domain. Moreover, it is possible to include the experimental measurements from sensors in the dataset. These advantages show the flexibility when preparing the data for DeepONets training.

\begin{figure}[ht]
\begin{center}
\includegraphics[width=\textwidth]{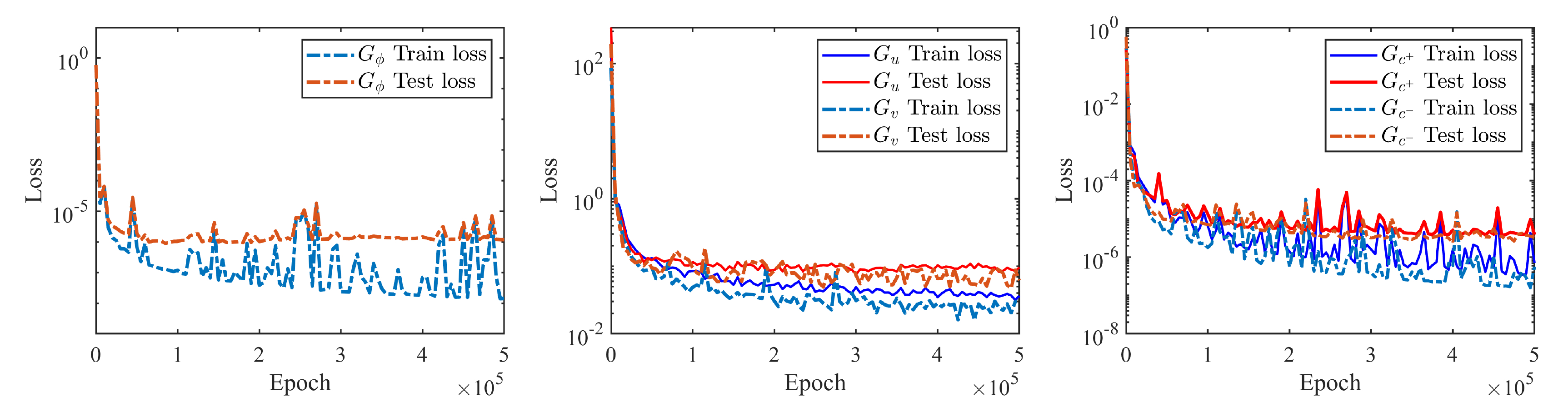}
\caption{DeepONets for 2D electroconvection: training and testing losses of $G_{\phi,u,v,c^{+},c^{-}}$. Note that we use MSE loss for training $G_{\phi,c^{+},c^{-}}$ and MAPE loss for $G_{u,v}$, thus the magnitudes of the loss functions are different. 
}\label{fig:EC2D_DeepONets_losses}
\end{center}
\end{figure}

To train the DeepONets, we apply the Adam optimizer with a small learning rate $2\times10^{-4}$, and the networks are trained over 500,000 iterations. The activation function of the neural network is ReLU. The losses of the training data and testing data during training process are presented in Figure \ref{fig:EC2D_DeepONets_losses}. As shown, the losses converge to small values. Note that we apply MAPE loss function to $G_{u,v}$, and thus the magnitude of their losses is different from the others. Upon training, these DeepONets can predict all fields accurately when the input functions are given. 

\subsection{Testing of DeepONets}

\begin{figure}[ht]
\begin{center}
\includegraphics[width=\textwidth]{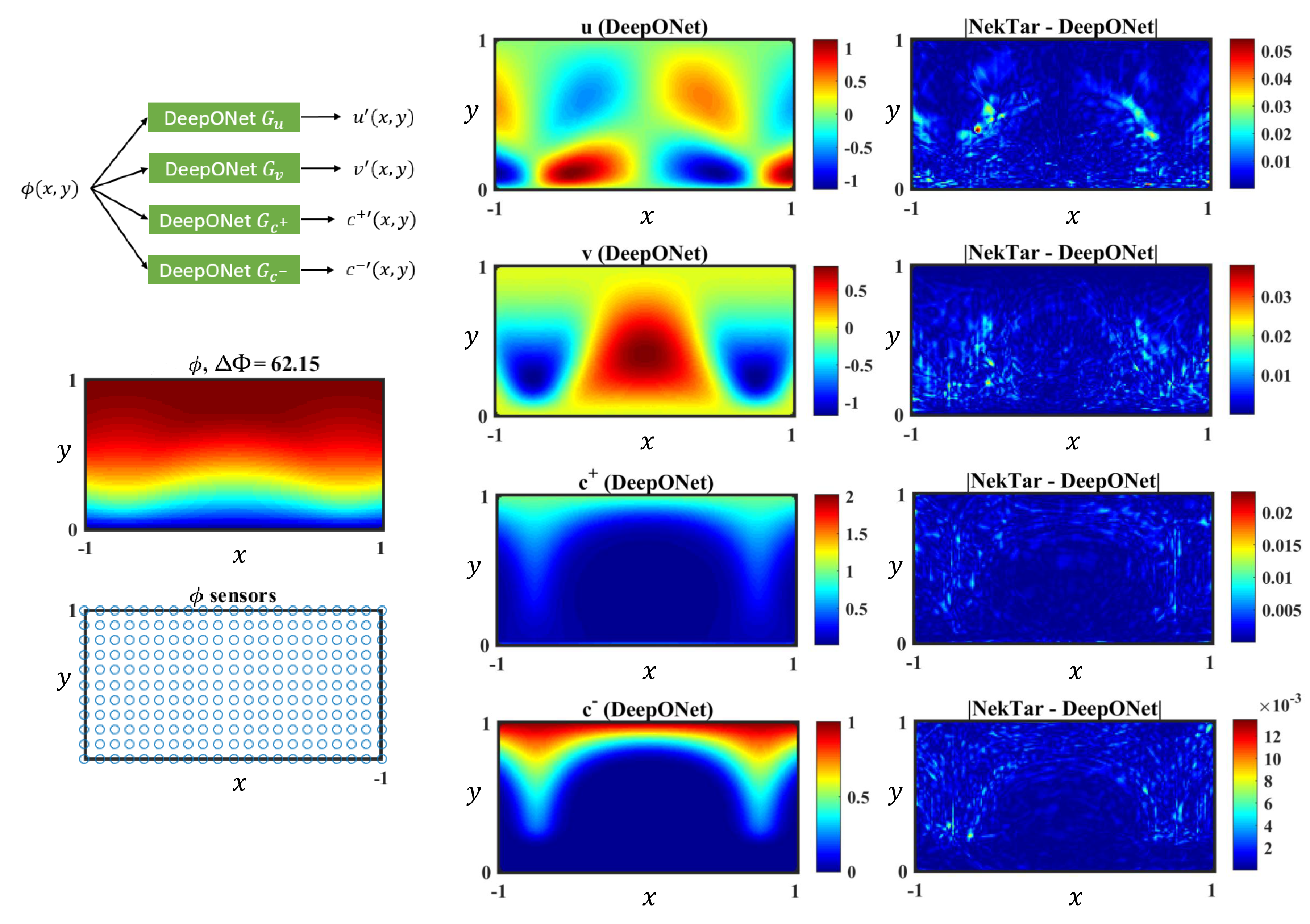}
\caption{DeepONets (upper left) for 2D electroconvection: predictions of $u$, $v$, $c^{+}$ and $c^{-}$ (from top to bottom) using DeepONets $G_{u}$, $G_{v}$, $G_{c^{+}}$ and $G_{c^{-}}$ for a new case 
corresponding to $\Delta \Phi=62.15$. Note that $\phi(x,y)$ is the input of these networks, and $u$, $v$, $c^{+}$ and $c^{-}$ are the outputs of four independent DeepONets. For comparison, we generate the reference fields by using NekTar simulations and plot the absolute errors (right column). 
}\label{fig:EC2D_DeepONets_pred1}
\end{center}
\end{figure}

\begin{figure}[h]
\begin{center}
\includegraphics[width=\textwidth]{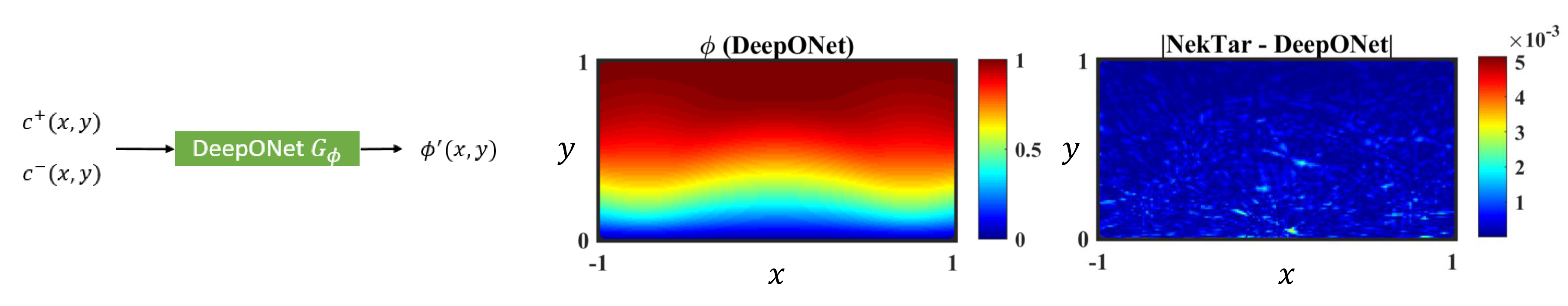}
\caption{A DeepONet for 2D electroconvection: predictions of $\phi$ using the DeepONet $G_{\phi}$ in the case of $\Delta \Phi=62.15$. Note that $c^{+}$ and $c^{-}$ are the inputs of this network, and $\phi(x,y)$ is the output. For comparison, we generate the reference field by using NekTar simulations and plot the absolute error (right plot). 
}\label{fig:EC2D_DeepONets_pred2}
\end{center}
\end{figure}

Here, we demonstrate the predictions of the well-trained DeepONets for the new case (not included in the training set) corresponding to  $\Delta \Phi=62.15$. 
The results of $G_{u}$, $G_{v}$, $G_{c^{+}}$ and $G_{c^{-}}$ are shown in Figure \ref{fig:EC2D_DeepONets_pred1}. In this figure, the input function $\phi(x,y)$, which is represented by a set of values at the sensor locations, is fed into the networks. We find that the 
DeepONet predictions are in good agreement with 
the fields generated by NekTar simulation. The MSEs for different state variables are on the order of $\mathcal{O}(10^{-6}-10^{-5})$. 
On the other hand, for testing the DeepONet $G_{\phi}$, the inputs $c^{+}$ and $c^{-}$ are provided. The prediction of $\phi$ is illustrated in Figure \ref{fig:EC2D_DeepONets_pred2}. 

The prediction errors for both $\Delta \Phi=13.4$ and $\Delta \Phi=62.15$ 
are given in Table \ref{tab:DeepONets_2D_errors}. In addition to MSEs, we also compute the relative $L_2$-norm errors, namely $\epsilon_{V} = \parallel V- \hat{V}  \parallel_{2} / \parallel V \parallel_{2} $. We find that in the case of $\Delta \Phi=13.4$, the $L_2$ errors of the velocity fields ($u$ and $v$) are much larger than the others. The reason is that the magnitude of the velocity fields is almost zero ($\mathcal{O}(10^{-4})$) 
in the case of $\Delta \Phi=13.4$. In fact, the predictions of DeepONets $G_{u}$ and $G_{v}$ are very close to the references, which can be seen from the small MSEs. 
To deal with the multiscale velocity, we apply the MAPE loss for $G_{u}$ and $G_{v}$. A comparison between the MSE and MAPE applied to $G_{u}$ is demonstrated in \ref{sec:app_MSE_VS_MAPE}.


We should note that once the DeepONets have been well trained, the prediction is a simple network evaluation task that can be performed very efficiently.
In the 2D electroconvection problem, the training of a DeepONet requires approximately 2 hours. However, 
the well-trained DeepONets take less than 1~second (on any common GPU) to predict the 2D fields when the input functions are provided. 
The speedup of DeepONets prediction versus the NekTar simulation for independent conditions is about 10,000 folds.

\begin{table}[h]
\begin{center}
\caption{DeepONets for 2D electroconvection: MSEs and relative $L_2$-norm errors of the DeepONet predictions for two testing examples. In the case of $\Delta \Phi=13.4$, the $L_2$ errors of the velocities ($u$ and $v$) are much larger than the others brcause the magnitude of the velocity in this case is almost zero ($\mathcal{O}(10^{-4})$). 
 }\label{tab:DeepONets_2D_errors}
\begin{tabular}{|c|cc|cc|}
  \hline
  \multirow{2}{*}{DeepONets} & \multicolumn{2}{c|}{$\Delta \Phi=13.4$}  & \multicolumn{2}{c|}{$\Delta \Phi=62.15$} \\
  & MSE & $L_2$ error & MSE & $L_2$ error  \\
  \hline
  $G_{\phi}$  &  $3.85\times10^{-6}$  & 0.23\%  & $1.53\times10^{-7}$   & 0.05\% \\
  $G_{u}$      &  $1.92\times10^{-9}$  & 8.54\%  & $3.89\times10^{-5}$   & 1.65\%  \\
  $G_{v}$      &  $2.55\times10^{-9}$  & 10.01\%  & $1.51\times10^{-5}$   & 0.86\% \\
  $G_{c^{+}}$   &  $6.59\times10^{-6}$  & 0.49\%  & $1.29\times10^{-6}$   & 0.31\% \\
  $G_{c^{-}}$  &  $4.20\times10^{-6}$  & 0.36\%  & $3.64\times10^{-6}$   & 0.45\%  \\
  \hline
\end{tabular}
\end{center}
\end{table}

\clearpage
\section{DeepM$\&$Mnet for 2D Electroconvection}\label{sec:DeepMM}

In order to use the pre-trained DeepONets, the input function  should be given. For example, the proper electric potential $\phi(x,y)$ should be provided to $G_{u}$ to obtain the $u$-velocity. 
However, this is not realistic in general. Therefore, we develop a new framework called DeepM$\&$Mnet, which allows us to infer the full fields of 
the coupled electroconvection problem when only a few measurements for any of the state variables are available. In the context of DeepM\&Mnet, a neural network is used to approximate the solutions of the electroconvection, while the pre-trained DeepONets are applied as the constraints of the solutions. In the following, we introduce two architectures of the proposed DeepM\&Mnet 
in \ref{sec:deepMM_architecture} and present the evaluation results in \ref{sec:deepMM_results}.  

\subsection{DeepM$\&$Mnet architectures}\label{sec:deepMM_architecture}

\subsubsection{Parallel DeepM$\&$Mnet archtecture}

A schematic diagram of the parallel DeepM\&Mnet architecture is illustrated in Figure \ref{fig:EC2D_DeepMM_parallel}. In this context, a fully-connected network with trainable parameters is used to approximate the coupling solutions. This is an ill-posed problem since only a few  measurements are available. Therefore, regularization is required to avoid overfitting. 
In this work, the pre-trained DeepONets are applied to deal with this problem. The pre-trained DeepONets are fixed (not trainable) and considered as the constraints of the NN outputs, which can be seen in Figure \ref{fig:EC2D_DeepMM_parallel}. The neural network, which takes $(x,y)$ as inputs and outputs $(\phi, u, v, c^{+}, c^{-})$, is trained by minimizing the following loss function: 
\begin{equation}
    \begin{aligned}
        {\arg \min}_{\theta} ~ \mathcal{L} = \lambda_{d}\mathcal{L}_{data} + \lambda_{o}\mathcal{L}_{op} + \lambda_{r}\mathcal{L}_{2}(\theta) ,
    \end{aligned}
    \label{eq:EC2D_deepMM_parallel_Loss}
\end{equation}
where $\lambda_{d}$, $\lambda_{o}$ and $\lambda_{r}$ are the weighting coefficients of the loss function; $\mathcal{L}_{2}(\theta) = \parallel \theta \parallel_{2}$ is the $L_{2}$ regulatization of the trainable parameters $\theta$, which can help avoid overfitting and stabilize the training process\footnote{The influence of the $L_{2}$ regulatization term for DeepM\&Mnet is investigated in \ref{sec:app_l2reg}. }; and
\begin{equation*}
    \begin{aligned}
    \mathcal{L}_{data} &= \sum_{V\in{(\phi, u, v, c^{+}, c^{-})}} \frac{1}{N_{d}} \sum_{i=1}^{N_{d}} (V(x^{i},y^{i})-V_{data}(x^{i},y^{i})),  \\
    \mathcal{L}_{op} &=  \sum_{V\in{(\phi, u, v, c^{+}, c^{-})}} \frac{1}{N_{op}} \sum_{i=1}^{N_{op}} (V(x^{i},y^{i})-V'(x^{i},y^{i})),   
    \end{aligned}
\end{equation*}
where $\mathcal{L}_{data}$ is the data mismatch and $\mathcal{L}_{op}$ is the difference between the neural network outputs and the DeepONets outputs. $V$ can be any variables of the investigated solutions ($\phi, u, v, c^{+}, c^{-}$); $V(x^{i},y^{i}$) accounts for the output of the fully-connected network, while $V'(x^{i},y^{i})$ is the output of DeepONet. $N_{d}$ and $N_{op}$ denote the number of measurements 
for each variable and the number of points for evaluating the operators, respectively.  
Here, we would like to add some comments on DeepM\&Mnet. First, it is not necessary to have measurements for every variable. For example, if measurements are only available for $\phi$, the DeepONets $G_{u}$, $G_{v}$, $G_{c^{+}}$ and $G_{c^{-}}$ can provide constraints for the other variables and guide the NN outputs to the correct solutions. 
Second, in the framework of parallel DeepM\&Mnet (Figure \ref{fig:EC2D_DeepMM_parallel}), we do not only have the outputs of the fully-connected network $V\in \{\phi,u,v,c^{+},c^{-}\}$, but also the outputs of DeepONets $V'\in \{\phi',u',v',c^{+'},c^{-'}\}$. Ideally, $V$ and $V'$ should converge to the same values. However, there is bias between $V$ and $V'$ due to the approximation error of the pre-trained DeepONets and the optimization error of the neural network.


\subsubsection{Series DeepM$\&$Mnet architecture}

\begin{figure}[t]
\begin{center}
\includegraphics[width=\textwidth]{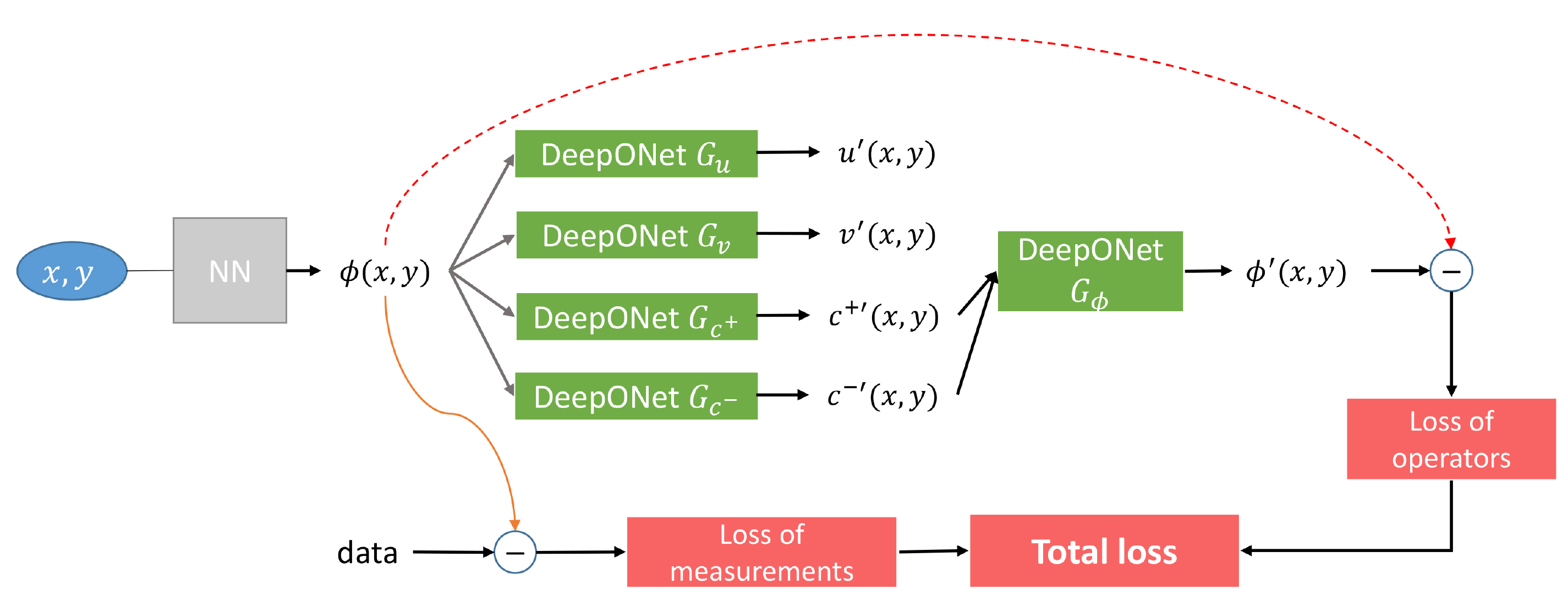}
\caption{DeepM\&Mnet for 2D electroconvection problem: schematic of the series architecture. Here, the neural network (NN), which is used to approximate only one solution (i.e., $\phi$), is a fully-connected network with trainable parameters. The other variables (i.e., $u, v, c^{+}, c^{-}$) are hidden outputs in this framework. 
The pretrained DeepONets are fixed (not trainable) and they are considered as the constraints of the NN outputs. }
\label{fig:EC2D_DeepMM_series}
\end{center}
\end{figure}

In addition to the parallel architecture, we also propose a series architecture of DeepM\&Mnet, which is illustrated in Figure \ref{fig:EC2D_DeepMM_series}. In the series setup, the fully-connected network is only used to approximate $\phi$. The other variables (i.e., $u, v, c^{+}, c^{-}$) are the hidden outputs in this framework and given by the DeepONets based on the result of $\phi$. Moreover, with the pre-trained DeepONet $G_{\phi}$, we can generate $\phi '$ from $(c^{+'}, c^{-'})$. 
The loss function is similar to (\ref{eq:EC2D_deepMM_parallel_Loss}).
However, here we assume that we only have the measurements of $\phi$, and $\mathcal{L}_{op}$ only contains the $\phi$ operator, thus:
\begin{equation*}
    \begin{aligned}
    \mathcal{L}_{data} &=  \frac{1}{N_{d}} \sum_{i=1}^{N_{d}} (\phi(x^{i},y^{i})-\phi_{data}(x^{i},y^{i})),  \\
    \mathcal{L}_{op} &=  \frac{1}{N_{op}} \sum_{i=1}^{N_{op}} (\phi(x^{i},y^{i})-\phi'(x^{i},y^{i})).   
    \end{aligned}
\end{equation*}
Different from the parallel architecture, in the series DeepM\&Mnet, $V$ only represents $\phi$, while $V'\in \{\phi',u',v',c^{+'},c^{-'}\}$. 
This framework shows that given a few measurements of $\phi$, the neural network can produce the full field of $\phi$. All other fields can be obtained by the pre-trained DeepONets inside the loop.

\subsection{DeepM$\&$Mnet testing results}\label{sec:deepMM_results}

\begin{figure}[t]
\begin{center}
\includegraphics[width=\textwidth]{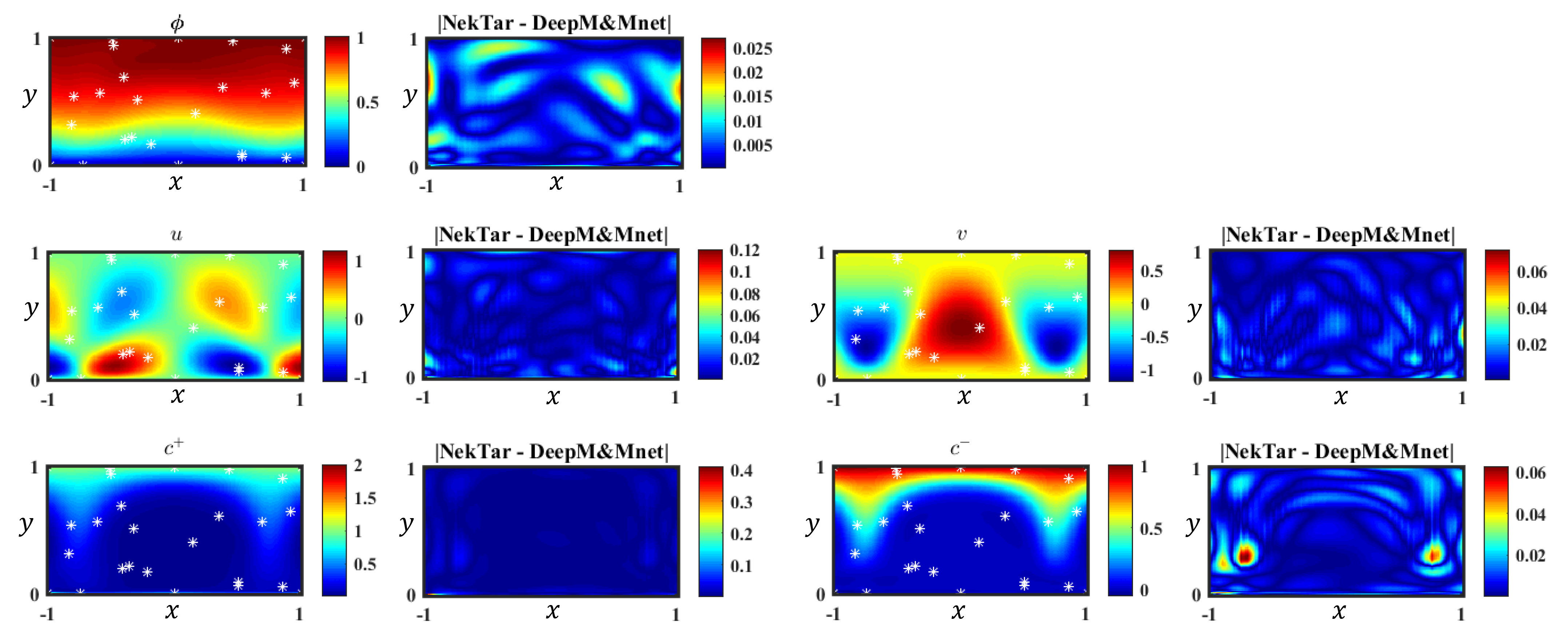}
\caption{DeepM\&Mnet for 2D electroconvection problem: results of parallel architecture. In this case, 20 measurements of each field are provided, which are marked with $*$. Left: results of the NN outputs (i.e., $\phi, u, v, c^{+}, c^{-}$), right: point-wise difference. 
}
\label{fig:EC2D_DeepMM_parallel_results}
\end{center}
\end{figure}

We first investigate the performance of DeepM$\&$Mnet for the 2D electroconvection problem, where the data is generated by NekTar with $\Delta \Phi=62.15$. This corresponds to an input not seen by DeepM\&M nor by the DeepONets, which are its building blocks. First, 20 measurements of each field are provided for the parallel architecture. The neural network contains 6 hidden layers and 100 neurons per layer. The hyperbolic tangent function is applied 
as the activation function. 
The neural network is trained by the Adam optimizer for 60,000 iterations with a learning rate $5\times10^{-4}$. For the loss function, we set $\lambda_{d}=10$, $\lambda_{o}=1$ and $\lambda_{r}=10^{-4}$. 
To evaluate the operator loss, we randomly choose 1000 points inside the domain, namely $N_{op}=1000$.
The resulting fields of one typical training process are demonstrated in Figure \ref{fig:EC2D_DeepMM_parallel_results}. Point-wise differences between the NN outputs and the NekTar simulations are also given. The $L_2$-norm errors of $V$ are: $\epsilon_{\phi}=0.80\%, \epsilon_{u}=3.14\%, \epsilon_{v}=1.85\%, \epsilon_{c^{+}}=3.24\%, \epsilon_{c^{-}}=2.97\%$. 
In addition, we can obtain the solutions from the DeepONet outputs $V'$. The $L_2$-norm errors of $V'$ are: $\epsilon_{\phi'}=0.58\%, \epsilon_{u'}=1.34\%, \epsilon_{v'}=0.70\%, \epsilon_{c^{+'}}=0.68\%, \epsilon_{c^{-'}}=0.78\%$. 

\begin{figure}[t]
\begin{center}
\includegraphics[width=\textwidth]{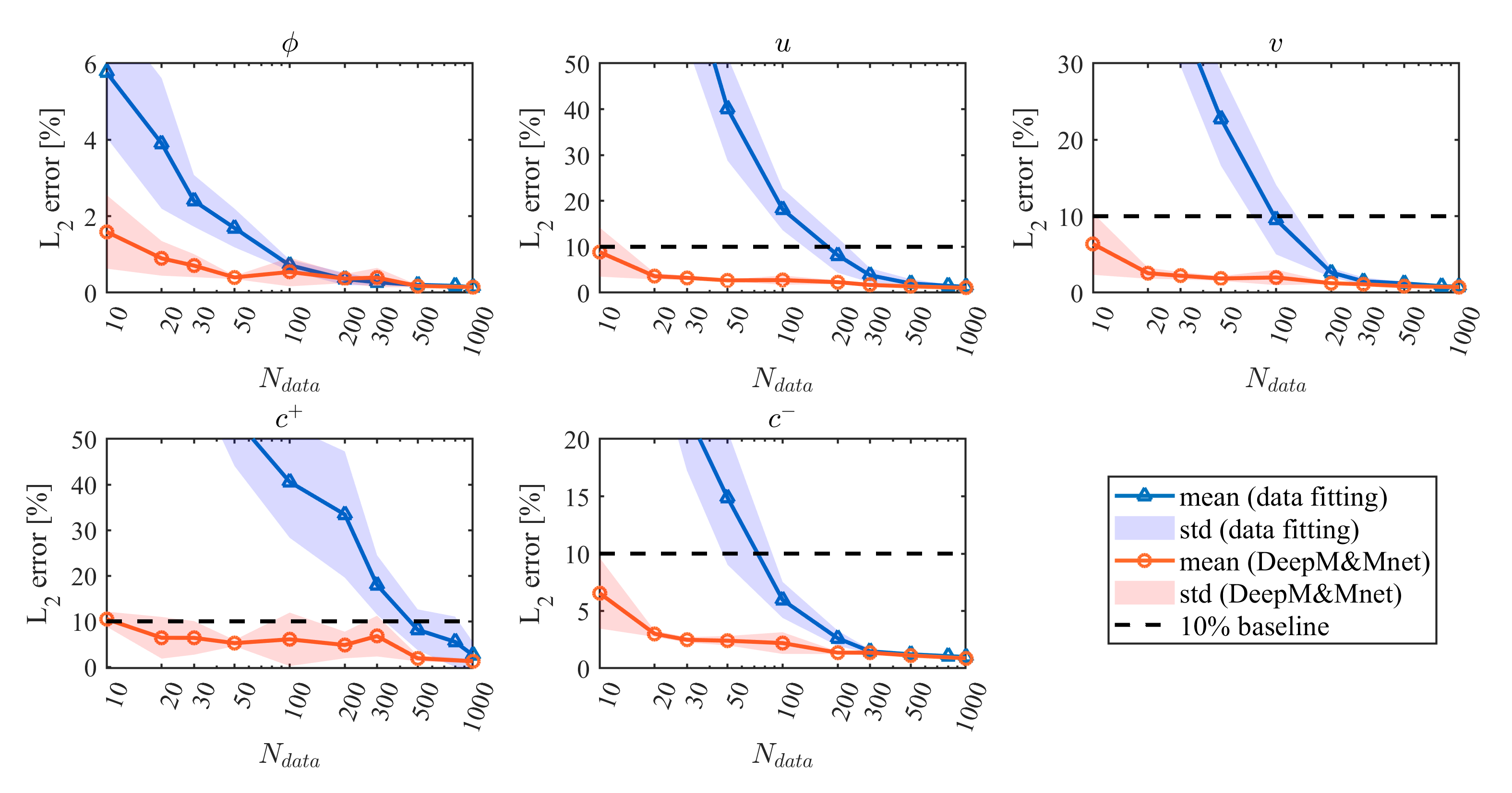}
\caption{DeepM\&Mnet for 2D electroconvection problem: relative $L_2$-norm errors with respect to the number of measurements ($N_{data}$). The DeepM\&Mnet results are from the neural network (namely $V\in\{\phi,u,v,c^{+},c^{-}\}$) of the parallel architecture. For each setting, 5 independent training processes with randomly-distributed sensors are applied. The mean errors and the standard deviations are computed. 
}
\label{fig:EC2D_DeepMM_sysNdata}
\end{center}
\end{figure}

Next, we also investigate the effect of the number of sensors ($N_{data}$). Figure \ref{fig:EC2D_DeepMM_sysNdata} shows the convergence of the relative $L_2$-norm errors with respect to the number of measurements. For comparison, we perform data fitting using a standard neural network for the same problem, in which the operator loss $\mathcal{L}_{op}$ is not considered, namely $\mathcal{L} = \lambda_{d}\mathcal{L}_{data} +  \lambda_{r}\mathcal{L}_{2}(\theta)$. 
We can see from the figure that without the DeepONet constraints, the neural 
network requires hundreds of sensors to obtain good accuracy (less than 10\%) for every state variable. However, only 10 measurements of each field are sufficient for DeepM\&Mnet to achieve the same accuracy. 
Note that the sensors are randomly distributed inside the domain. Therefore, we perform 5 independent training processes with different sensor locations, and then compute the mean errors and standard deviations. 
In this assessment, we use sensors for all the state variables. However, similar performance of DeepM\&Mnet can be obtained if we only have the data measurements of $\phi$. The reason is that those pre-trained DeepONets $G_{u,v,c^{+},c^{-}}$ can provide strong constraints on the other variables.

\begin{figure}[t]
\begin{center}
\includegraphics[width=0.9\textwidth]{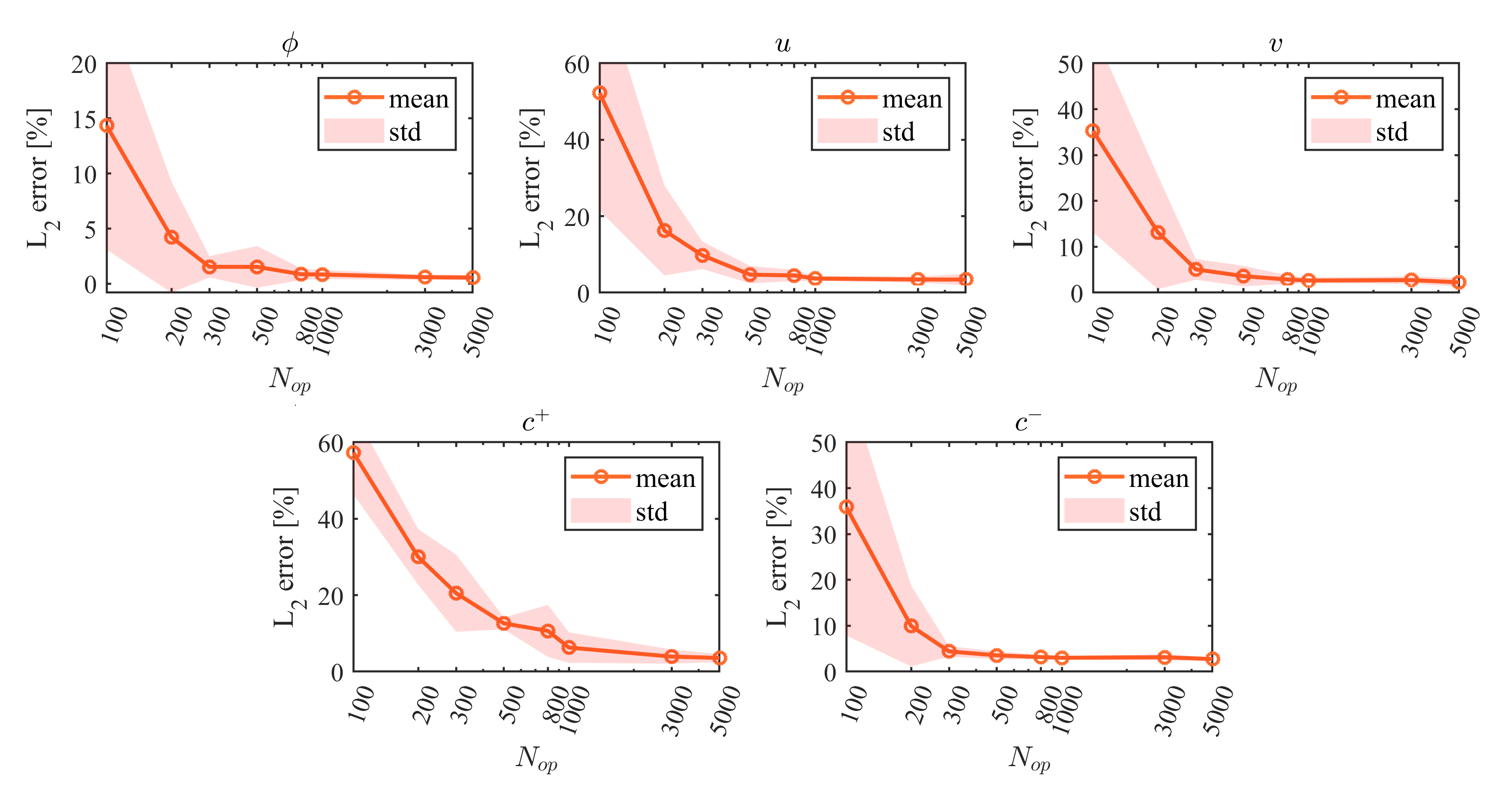}
\caption{DeepM\&Mnet for 2D electroconvection problem: relative $L_2$-norm errors with respect to the number of operator evaluation points ($N_{op}$). The DeepM\&Mnet results are from the neural network (namely $V\in\{\phi,u,v,c^{+},c^{-}\}$) of the parallel architecture. For each setting, 5 independent training processes with randomly-distributed sensors are applied. The mean errors and the standard deviations are computed. 
}
\label{fig:EC2D_DeepMM_sysNop}
\end{center}
\end{figure}

Moreover, we also demonstrate the influence of the number of operator evaluation points, namely $N_{op}$. 
We use 20 measurements for each condition. 
The relative $L_2$-norm errors are demonstrated in Figure \ref{fig:EC2D_DeepMM_sysNop}. We find that with more evaluation points for DeepONets the results are more accurate, since there are more constraints on the solutions (NN outputs).  Similarly, 5 independent training processes are applied and the mean errors and standard deviations are given in the plots.

\begin{figure}[t]
\begin{center}
\includegraphics[width=\textwidth]{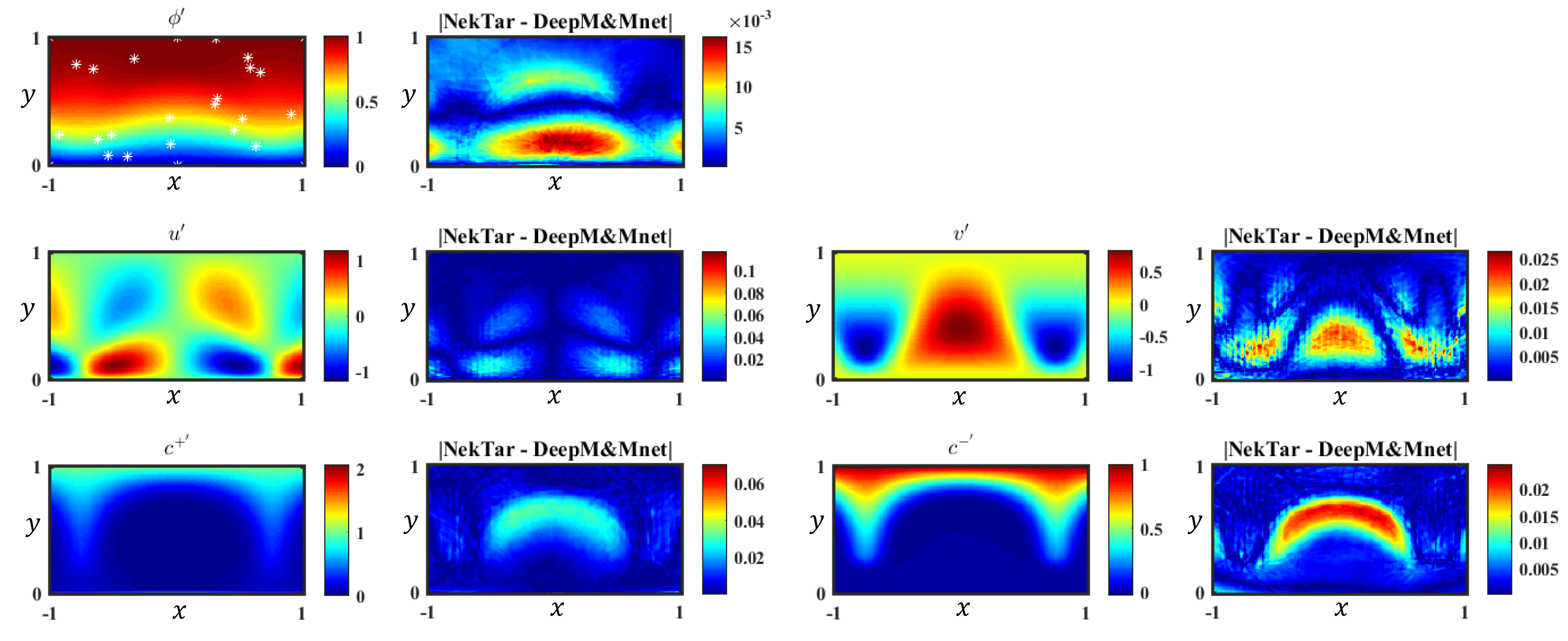}
\caption{DeepM\&Mnet for 2D electroconvection problem: results of {\em series} architecture. In this case, 20 measurements of $\phi$ are provided, which are marked with $*$. Left: results of the DeepONets outputs (i.e., $\phi', u', v', c^{+'}, c^{-'}$), right: point-wise difference. 
}
\label{fig:EC2D_DeepMM_series_results}
\end{center}
\end{figure}

For testing the {\em series architecture} of DeepM\&Mnet, only the data of $\phi$ is given. An example of the resulting fields is shown in Figure \ref{fig:EC2D_DeepMM_series_results}. In this case, 20 measurements of $\phi$ are provided and $N_{op}=1000$. The results of the other fields can be generated from the pre-trained DeepONets. As shown in the figure, the outputs are very consistent with the simulation. The relative The $L_2$-norm errors are: $\epsilon_{\phi'}=0.76\%, \epsilon_{u'}=3.75\%, \epsilon_{v'}=1.72\%, \epsilon_{c^{+'}}=2.70\%, \epsilon_{c^{-'}}=1.96\%$.

\clearpage
The results shown above indicate that DeepM\&Mnets can infer full fields of electroconvection accurately with a very small number of measurements. For conventional numerical methods, in which the physical model is generally required, it is difficult and time-consuming to solve such assimilation problem. 
On the contrary, the proposed DeepM\&Mnet framework is flexible since the DeepONets are pre-trained offline and are used in a ``plug-and-play" mode. It is also efficient as the DeepONets are fixed and only a simple NN is required to train. In this paper, the training of a DeepM\&Mnet takes about 20-30 minutes on a single GPU (NVIDIA Quadro RTX 6000).

\section{Summary}\label{sec:conclusion} %

In this paper, we first demonstrate the effectiveness of DeepONets for complex fluid systems using the multiphysics of electroconvection as a benchmark. We train several DeepONets to predict one of the coupled fields from the rest of the fields.
In our example, we use 15 different conditions corresponding to a range of applied electric potentials to form the training dataset. Upon training, the DeepONets can predict all fields very accurately and efficiently for any condition when the input is given. 

More importantly, in this paper, we propose a novel framework, the DeepM\&Mnet, which allows to integrate the pre-trained DeepONets and a few measurements from any of the fields, to produce the full fields of the coupled system. In DeepM\&Mnets, we use a neural network as the surrogate model of the multiphysics solutions, and use the 
pre-trained DeepONets as the constraints for the solutions. For both {\em parallel} and {\em series} DeepM\&Mnets, we find that only a few measurements are sufficient to infer the full fields of the electroconvection, even if measurements are not available for all state variables. 
The DeepM\&Mnet, which can be considered as a simple data assimilation framework, is much more flexible and efficient than any other conventional numerical method in terms of dealing with such assimilation problem. 
Note that in order to use the DeepM\&Mnets, the building blocks - the DeepONets - are required to be pre-trained with labeled data. However, as shown in the paper, preparing the training data is very flexible and the training can be done offline. Once the DeepONets have been trained and embedded in the the DeepM\&Mnet, it is straightforward to predict the solutions of a complex multiphysics and multiscale system when only a few measurements are available. More broadly, the results in this paper show that the new framework can be used for any type of multiphysics and multiscale problems, for example in hypersonics with finite chemistry, and we will report such results in future work.

\section*{Acknowledgements} %

The authors acknowledge support from DARPA/CompMods HR00112090062 and DOE/PhILMs 
DE-SC0019453.


\appendix
\setcounter{table}{0}
\setcounter{figure}{0}

\section{DeepONet training: MSE loss vs. MAPE loss}\label{sec:app_MSE_VS_MAPE}

As mentioned in Section \ref{sec:DeepONets}, we apply the MAPE loss to the DeepONets training for $G_{u,v}$ due to the large range of the velocity magnitude in the electroconvection problem. 
If we use a MSE loss for such multiscale prediction, the training loss and the back-propagation process will be dominated by those velocities with larger magnitudes, as the small-scale velocities contribute little to the NN optimization. Therefore, the MAPE loss is better for the DeepONets that are required to predict multiscale values. 

\begin{figure}[t]
\begin{center}
\includegraphics[width=\textwidth]{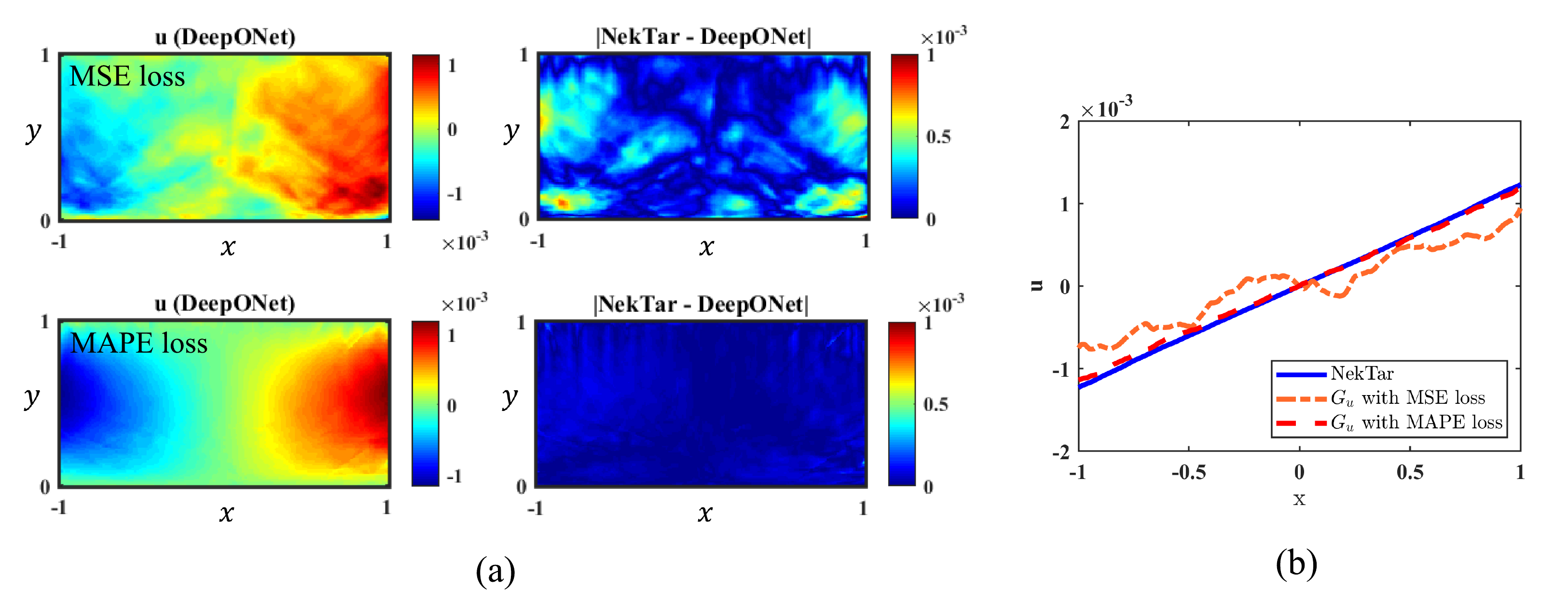}
\caption{Predictions of $G_{u}$ with different losses: (a) 2D fields and errors of $u$-velocity (the first row is $G_{u}$ with MSE loss and the second is $G_{u}$ with MAPE loss); (b) 1D profiles at $y=0.5$.
The testing example shown here is the case when $\Delta\Phi=13.4$.
}
\label{fig:EC2D_DeepONets_MSEvsMAPE_134}
\end{center}
\end{figure}

A comparison between the MSE and MAPE losses applied to $G_{u}$ is shown here. We learn from Figure \ref{fig:EC2D_DeepONets_dataDemo} that the $u$-velocity is a multiscale state variable. In particular, the magnitude of the velocity is approximately zero ($10^{-4}$) when $\Delta\Phi$ is small (e.g., $\Delta\Phi=13.4$), while it is in the order of $10^{0}$ when $\Delta\Phi$ is large (e.g., $\Delta\Phi=62.15$). 
The prediction errors of $G_{u}$ with MSE loss and MAPE loss are given in Table \ref{tab:DeepONets_Gu_MSE_MAPE}. 
We can see that the $G_{u}$ with MAPE loss performs much better than the one with MSE loss for $\Delta\Phi=13.4$. Although for $\Delta\Phi=62.15$, for $G_{u}$ with MAPE loss we get larger errors, it is still very accurate as the relative $L_2$-norm error is only $1.65\%$. 
To demonstrate the superiority of the MAPE loss for $G_{u}$, we show the predicted fields with different losses in Figure \ref{fig:EC2D_DeepONets_MSEvsMAPE_134}. The DeepONet with MSE loss fails to predict the flow pattern at $\Delta\Phi=13.4$ accurately, while the prediction of $G_{u}$ with MAPE loss is very consistent with the NekTar high-fidelity simulation.

\begin{table}[th]
\begin{center}
\caption{Prediction errors of $G_{u}$ with MSE loss and MAPE loss. 
 }\label{tab:DeepONets_Gu_MSE_MAPE}
\begin{tabular}{|c|cc|cc|}
  \hline
  \multirow{2}{*}{$G_{u}$} & \multicolumn{2}{c|}{$\Delta \Phi=13.4$}  & \multicolumn{2}{c|}{$\Delta \Phi=62.15$} \\
  & MSE & $L_2$ error & MSE & $L_2$ error  \\
  \hline
  MSE loss  &  $5.41\times10^{-8}$  & 45.27\%  & $1.10\times10^{-5}$   & 0.88\% \\
  MAPE loss      &  $1.92\times10^{-9}$  & 8.54\%  & $3.89\times10^{-5}$   & 1.65\%  \\
  \hline
\end{tabular}
\end{center}
\end{table}

\section{Effect of $L_2$ regularization for DeepM$\&$Mnet}\label{sec:app_l2reg}

The loss function of DeepM\&Mnet training, given in Equ. (\ref{eq:EC2D_deepMM_parallel_Loss}), is composed of a data term, an 
operator term and a $L_{2}$ regularization term for the trainable parameters. The $L_{2}$ regularization plays an important role in avoiding overfitting for data-friven methods. In this section, we present an example showing the effect of the $L_{2}$ regularization. 

We use the 1D electroconvection as the benchmark here. For 1D electroconvection, the electric potential $\phi$
and the concentration $c^{-}$ are functions of $y$. We train two independent DeepONets $G_{\phi}$ and $G_{c^{-}}$, where $G_{\phi}$ takes $c^{-}(y)$ as input and outputs $\phi(y)$, while $G_{c^{-}}$ performs oppositely. We generate 24 training trajectories for the 1D electroconvection at various values of $\Delta\Phi$, namely $\Delta\Phi=5, 10, \dots, 120$. For the individual DeepONet, two hidden layers and 100 neurons per layer are applied for both branch net and trunk net. Upon training, the DeepONets $G_{\phi}$ and $G_{c^{-}}$ can predict 
one field from another very accurately. 

\begin{figure}[ht]
\begin{center}
\includegraphics[width=0.9\textwidth]{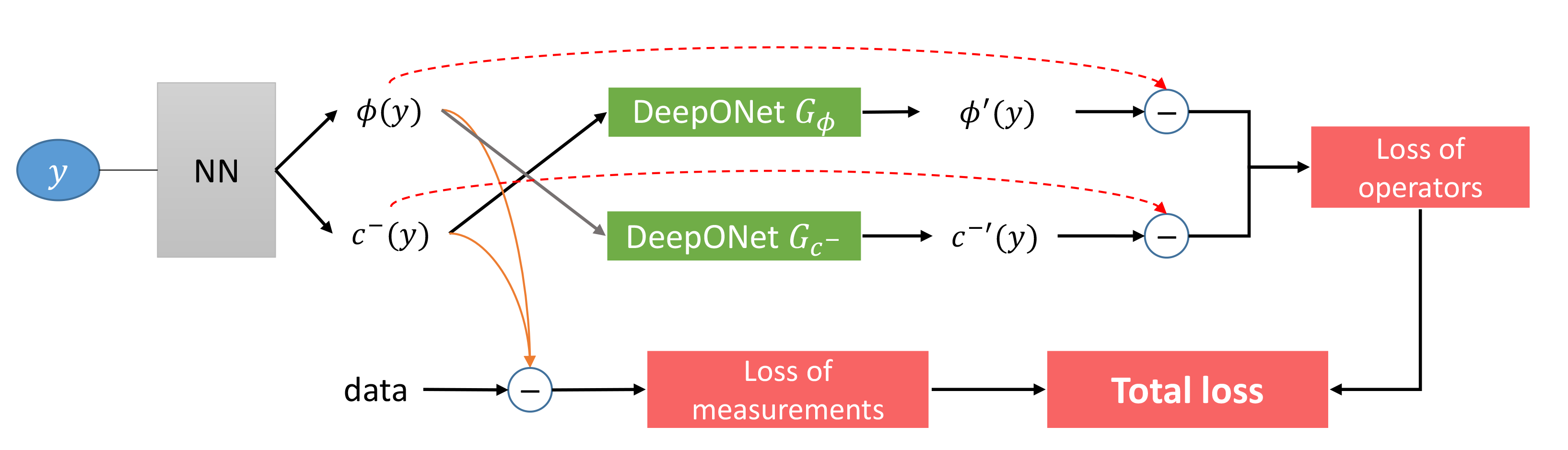}
\caption{DeepM\&Mnet for 1D electroconvection problem: schematic of the {\em parallel} architecture. Here, the neural network (NN) is used to approximate the 1D coupled solutions (i.e., $\phi$ and $c^{-}$). 
The pretrained DeepONets $G_{\phi}$ and $G_{c^{-}}$ are fixed and form constraints for the NN outputs. }
\label{fig:EC1D_DeepMM_parallel}
\end{center}
\end{figure}

\begin{figure}[ht]
\begin{center}
\includegraphics[width=\textwidth]{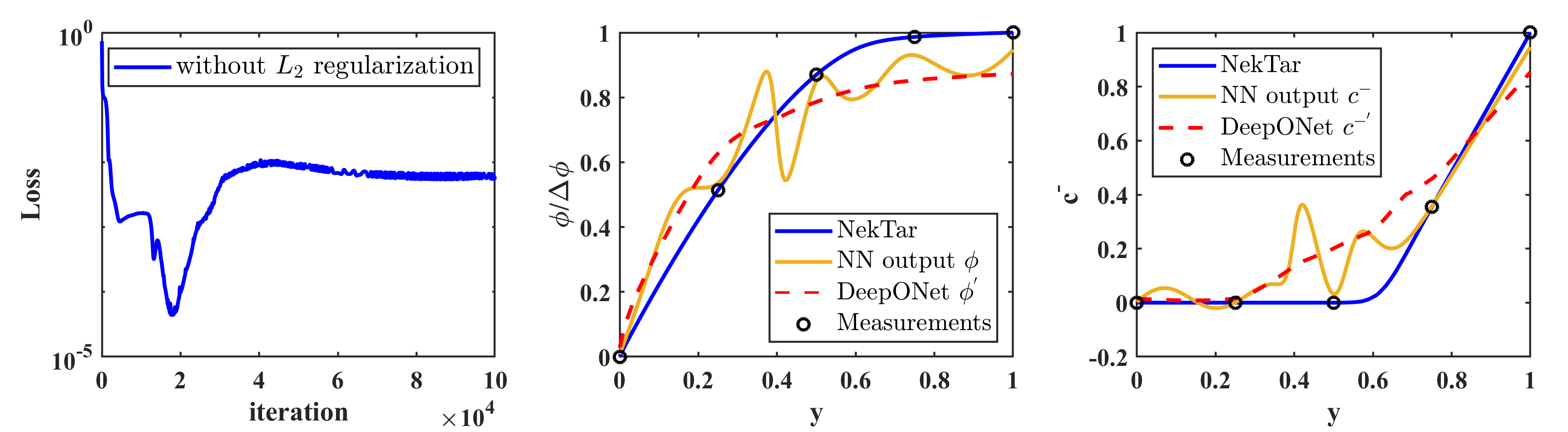}
\includegraphics[width=\textwidth]{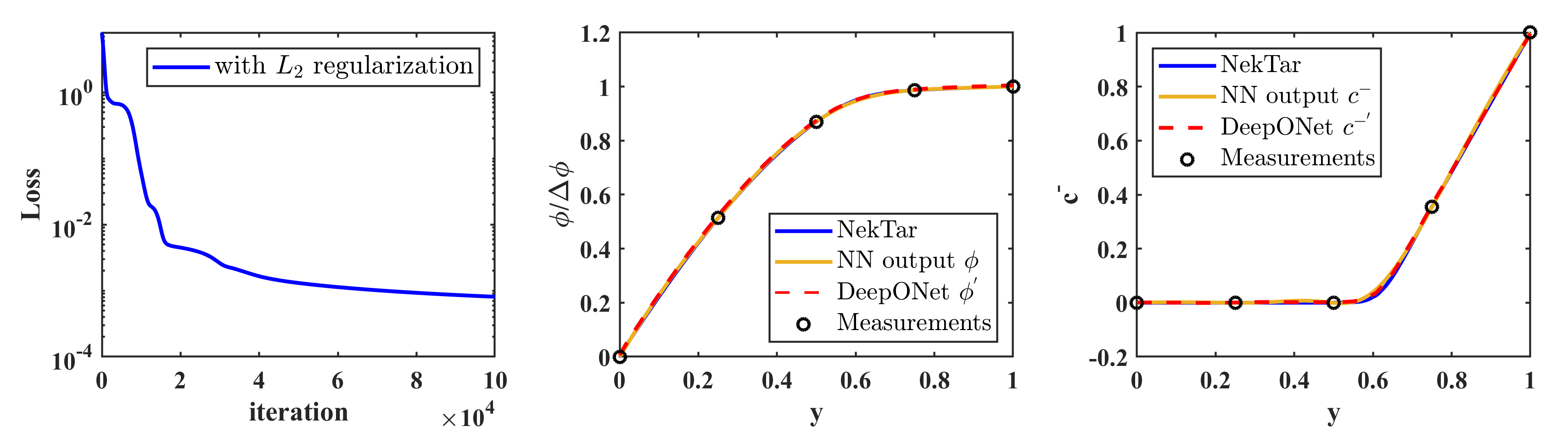}
\caption{Parallel DeepM\&Mnet for 1D electroconvection problem: results with different loss functions for $\Delta\Phi=77.6$. First row: loss without $L_2$ regularization; second row: loss with $L_2$ regularization. Left: training loss; middle: solution of $\phi(y)$; right: solution of $c^{-}(y)$. 
Here, 5 measurements are used for each coupled field (i.e., $\phi$ and $c^{-}$). Overfitting is observed in the case without $L_2$ regularization. 
}
\label{fig:EC1D_DeepMM_with_without_L2}
\end{center}
\end{figure}

Finally, we can construct the parallel DeepM\&Mnet architecture, which is illustrated in Figure \ref{fig:EC1D_DeepMM_parallel}. A neural network is used to approximate the solutions of $\phi(y)$ and $c^{-}(y)$. 
We apply the proposed DeepM\&Mnet to infer the profiles in the case of $\Delta\Phi=77.6$. The results with different loss functions are demonstrated in Figure \ref{fig:EC1D_DeepMM_with_without_L2}. Five measurements, which are uniformly distributed along $y$ direction, are used for each coupled field. We find that in the case where the $L_{2}$ regularization is not considered, the loss diverges and the solutions are overfitting to the measurements. 
This phenomenon is observed frequently when we perform ensemble simulations. 
On the contrary, the DeepM\&Mnet with  $L_{2}$ regularization in the loss function is much more robust and attains consistent results with good accuracy for every simulation.

\bibliographystyle{elsarticle-num-names}
\bibliography{ref.bib}







\end{document}